\documentclass[1p,authoryear]{elsarticle}
 
\usepackage{epsfig,lineno}
\usepackage{amsmath}
\usepackage{graphicx,subcaption,caption}
\usepackage{natbib}
\modulolinenumbers[1]
 
\def\tnit{\hbox{T$_{\hbox{N}_2}$}}
\def\pnit{\hbox{P$_{\hbox{N}_2}$}}
\def\znit{\hbox{Z$_{\hbox{N}_2}$}}

\def\anit{\hbox{N$_2^\alpha$}}
\def\bnit{\hbox{N$_2^\beta$}}
\def\nit{\hbox{N$_2$}}
\def\met{\hbox{CH$_4$}}
\def\wat{\hbox{H$_2$O}}
\def\mub{\hbox{\,$\mu$bar}}
\def\micron{\hbox{\,$\mu$m}}
\def\mum{\hbox{\,$\mu$m}}
\def\eg{\hbox{\em e.g.}}  
\def\ie{\hbox{\em i.e.}}   
\def\etal{\hbox{\em et al.}}
\def\deg{\hbox{$^\circ$}}
\def\tiu{\hbox{J\,m$^{-2}$\,s$^{-1/2}$\,K$^{-1}$}}
 
\begin{document}

\begin{frontmatter}

\title{Distribution and Energy Balance of Pluto's Nitrogen Ice, as seen by New Horizons in 2015}

\address[colum]{Department of Astronomy, Columbia University, 550 W. 120th St., New York, NY 10027}
\address[ucla]{Division of Astronomy and Astrophysics, University of California, Los Angeles, 475 Portola Plaza, Los Angeles, CA 90025}
\address[stsci]{Space Telescope Science Institute, 3700 San Martin Drive, Baltimore, MD 21218}
\address[low]{Lowell Observatory, 1400 W Mars Hill Rd, Flagstaff, AZ 86001}
\address[gren]{Universit\'e Grenoble Alpes,  France}
\address[umd]{Department of Astronomy, University of Maryland, College Park, College Park, MD 20742-2421}
\address[swri]{Southwest Research Institute,1050 Walnut St 300, Boulder, CO 80302}
\address[apl]{Johns Hopkins University Applied Physics Laboratory, 11100 Johns Hopkins Rd, Laurel, MD 20723}
\address[ames]{NASA Ames Research Center, Moffett Blvd, Mountain View, CA 94035}

\author[colum,ucla]{Briley L.~Lewis\corref{cor1}}
\ead{bll2124@columbia.edu}
\author[stsci]{John A. Stansberry\corref{cor2}}
\ead[js]{jstans@stsci.edu}
\author[stsci]{Bryan J.~Holler}
\author[low]{William M.~Grundy}  
\author[gren]{Bernard~Schmitt}
\author[umd]{Silvia~Protopapa}
\author[apl]{Carey~Lisse}
\author[swri]{S. Alan~Stern}
\author[swri]{Leslie~Young}
\author[apl]{Harold A.~Weaver}
\author[swri]{Catherine~Olkin}
\author[ames]{Kimberly~Ennico}
\author{and the New Horizons Science Team}

\cortext[cor2]{Corresponding author}
\cortext[cor1]{Primary corresponding author}
 

\begin{abstract}
Pluto's surface is geologically complex because of volatile ices that are mobile on seasonal and longer time scales. Here we analyzed New Horizons LEISA spectral data to globally map the nitrogen ice, including nitrogen with methane diluted in it. Our goal was to learn about the seasonal processes influencing ice redistribution, to calculate the globally averaged energy balance, and to place a lower limit on Pluto's \nit\ inventory. 
We present the average latitudinal distribution of nitrogen and investigate the relationship between its distribution and topography on Pluto by using maps that include the shifted bands of methane in solid solution with nitrogen (which are much stronger than the 2.15-\micron\ nitrogen band) to more completely map the distribution of the nitrogen ice. We find that the global average bolometric albedo is $0.83 \pm 0.11$, similar to that inferred for Triton, and that a significant fraction of Pluto's \nit\ is stored in Sputnik Planitia. We also used the encounter-hemisphere distribution of nitrogen ice to infer the latitudinal distribution of nitrogen over the rest of Pluto, allowing us to calculate the global energy balance. Under the assumption that Pluto's nitrogen-dominated 11.5\mub\ atmosphere is in vapor pressure equilibrium with the nitrogen ice, the ice temperature is $36.93 \pm 0.10$ K, as measured by New Horizons' REX instrument. Combined with our global energy balance calculation, this implies that the average bolometric emissivity of Pluto's nitrogen ice is probably in the range 0.47 -- 0.72. This is consistent with the low emissivities estimated for Triton based on Voyager results, and may have implications for Pluto's atmospheric seasonal variations, as discussed below. The global pattern of volatile transport at the time of the encounter was from north to south, and the transition between condensation and sublimation within Sputnik Planitia is correlated with changes in the grain size and \met\ concentration derived from the spectral maps. The low emissivity of Pluto's \nit\ ice suggests that Pluto's atmosphere may undergo an extended period of constant pressure even as Pluto recedes from the Sun in its orbit.
\end{abstract}


\end{frontmatter}
  
\section{Introduction}\label{intro}

Pluto's atmosphere was discovered in 1988 via stellar occultation  \citep{hubbard1988occultation,elliot1989plutoatmo}, and with the discovery of \nit\ ice on its surface, it became apparent that the atmosphere was likely dominated by \nit\ (not \met) in vapor pressure equilibrium with the ice (\eg\ \citealt{owen1993plutoatmo}). Due to the large latent heat of sublimation of \nit, sublimation and condensation efficiently re-distribute energy from highly-illuminated to less-illuminated (or dark) regions, with the result that the \nit\ ice everywhere is essentially isothermal (\eg\ \citealt{spencer1997volatile}, \citealt{trafton1983global}). This is also true for Triton's atmosphere (\eg\ \citealt{yelle1995tritonatmo}, \citealt{trafton1984}). However, because New Horizons included a mapping spectrometer (while Voyager did not) we now know in detail how the \nit\ ice is distributed on Pluto (but not on Triton). Thus we can relate the \nit\ (and CO and \met) ice distribution to the stunning diversity of terrain on Pluto's surface revealed by New Horizons in 2015. We can also use the measured distribution of \nit\ to calculate its energy balance, without the need to simply assume a relationship between albedo and composition as was necessary for Triton (\eg\ \citealt{stansberry1992tritonebal}). New Horizons also revealed that topography can strongly influence the distribution of volatile ices, namely that the Sputnik Planitia basin is filled with \nit, \met, and CO (as described in \citet{grundy2016surface}). That relationship was not understood based on Voyager observations of Triton (in part because of the very subdued topography there), but was predicted prior to the Pluto encounter \citep{stansberry2014plutotopo,trafton2015departure}, and further described post-encounter \citep{bertrand2016atmotopo}. 

Data from the Linear Etalon Imaging Spectral Array (LEISA) and Multi-spectral Visible Imaging Camera (MVIC) provided global views of composition, albedo, color and geology on Pluto \citep{reuter2008ralph, 2010Natur.468..775S}. In addition, the Long-Range Reconnaisance Imager (LORRI, \citet{cheng2008lorri}) provided albedo and limited geologic information at long ranges, particularly of Pluto's Charon-facing hemisphere, which was also not visible at New Horizons' closest approach. In this study, we used previously published maps created from LEISA data to create new global maps of  Pluto's nitrogen-ice deposits. We then combined these maps with existing albedo data and topographic maps to investigate correlations between nitrogen presence and insolation and topography, estimate the \nit\ ice bolometric albedo and emissivity, and describe how the \nit\ distribution varies with latitude.

Our goal was to use the New Horizons data sets to provide the most comprehensive map of \nit\ ice distribution on Pluto, relate that distribution to the geology (specifically topography), and precisely determine the global energy balance of that ice. These results provide a detailed snapshot of Pluto's seasonal and climatic state at the epoch of the New Horizons encounter. Future modeling work that investigates the seasonal evolution of the volatile ice distribution and atmospheric pressure and composition of Pluto can use our results as a starting point for predictions, or as a constraint informing historical studies. Previous studies have presented maps of the \nit\ ice distribution on the encounter hemisphere, based on the presence of the 2.15-\micron\ absorption band and $CH_4$ band shift, as well as maps for H$_2$0, CO and \met\ ices and tholins \citep{grundy2016surface,protopapa2017pluto,schmitt2017physical, 2010Natur.468..775S, 2015AGUFM.U53A..01S}, and for \met\ only \citep{earle2017long}. Some of these additional maps are used in our investigations of volatile abundance at various elevations in Section \ref{sec:elev}. Topographic data was also limited to the encounter ({\em i.e.} anti-Charon) hemisphere of Pluto, because it requires high resolution imagery, and is based on the digital elevation model (DEM) of 
\citet{schenk2018plutotopo}. We also relied on maps of bolometric albedo \citep{buratti2017global}, which cover both the encounter and sub-Charon hemispheres of Pluto, albeit at vastly different spatial resolution. All of these data products are limited to the illuminated latitudes of Pluto at the encounter epoch, {\em i.e.} northward of 38\deg S (except for some limited topographic information in the ``haze-illuminated'' region south of the terminator on the encounter hemisphere -- see \citealt{schenk2018plutotopo, 2010Natur.468..775S}). 

Another critical constraint is provided by the measurement of Pluto's atmospheric surface pressure. Because the \nit-dominated atmosphere and surface \nit\ ices are in effective vapor pressure equilibrium, the atmospheric surface pressure can be equated to the physical temperature (as distinguished from the brightness temperature, such as was measured using the REX instrument) of the \nit\ ice via the vapor pressure relation. \citet{hinson2017radio}, based on the REX radio-occultation results from ingress and egress, give a mean atmospheric surface pressure of $11.5 \pm\ 0.7$\mub, at a radius of 1189 km (within 1 km of Pluto's average radius, $1188.3 \pm 1.6$ km; \citealt{nimmo2017plutoradius}). Using the older vapor pressure curve of \citet{brown1980vappress}, this implies  a \nit-ice temperature of $37.18 \pm 0.10$ K. However, \citet{fray2009vappress} find that the vapor pressure over \nit\ ice is actually $\simeq 18$\% higher at these temperatures, implying a \nit-ice temperature of $36.93 \pm 0.10$ K. We adopt the \citet{fray2009vappress} results for the remainder of this paper. The small temperature error is a result of the steep dependence of the vapor pressure on temperature ({\em i.e.} of the large latent heat of sublimation of \nit). The REX instrument also obtained brightness temperature measurements for Pluto's near-surface at a wavelength of 4.2\,cm. Interpretation of those data in terms of the physical temperature of the surface, and specifically of areas covered in \nit\ ice, is ongoing, but seem to require a (non-bolometric) emissivity near 0.7 for Sputnik Planitia\footnote{Some place names on Pluto are a mix of informal and officially approved names.} (Linscott \etal, submitted.) We return to the topic of the emissivity of Pluto's \nit\ ice below.

Methane and \nit\ ices coexist as dilute solid solutions with one another. As reviewed by \citet{trafton2015phase},
the maximum \met\ and \nit\ mixing ratios in the other ice are 4\% and 3\%, respectively, at 37~K. Individual ice grains are thus either \nit\ dominated (\nit:\met) or \met\ dominated (\met:\nit). If the bulk composition of the ice in a particular location has a \met\ mixing ratio in the range 4\% -- 97\%, that ice is an assemblage of \met\ grains saturated in \nit\ and of \nit\ grains saturated in \met. The solution between these species will affect their vapor pressures, and is expected to result in non-ideal vapor pressure behavior (\ie\ not obeying Raoult's law: $P_{N_2}^{Sat} = X_{N_2} \times P_{N_2}^{Vap}$). Trafton (2015) shows that the vapor pressure of \nit\ over an assemblage of grains of saturated \nit:\met\ could be significantly lower than the Rauolt's law prediction. However, laboratory data constraining the vapor pressure behavior is lacking, so it is currently unknown how much the vapor pressure of \nit\ is depressed (nor how much the vapor pressure of \met\ over saturated \met:\nit\ is increased). There are now theoretical equations of state available, but they have yet to be validated by data, and thus were not included in this analysis. Lacking specific information on these effects, we assumed that Pluto's \nit:\met\ ice behaves very much like \nit\ ice, and \met:\nit\ behaves very much like \met\ ice, in terms of vapor pressure (with application to determining the surface pressure of the \nit-dominated atmosphere) and sublimation and condensation (with application to volatile transport rates and global redistribution of insolation via latent heat transport).

In Section \ref{methods}, we explain the methods used to create new maps of the distribution of \nit\ ice needed for this investigation, and how elevation relates to the distribution of this volatile. 
Then, in Section \ref{energy}, we calculate the emissivity required to produce an ice temperature of 37 K using principles of energy balance, utilizing our global nitrogen maps, bolometric albedo maps based on LORRI data \citep{cheng2008lorri} and the results of \cite{buratti2017global}, and assumed \nit-ice scenarios for the unobserved South Pole. Lastly, Section \ref{results} explores the possible physical causes of the observed trends and the implications of our energy balance calculations, drawing on other long-term volatile transp.pngort and climate modeling efforts from the New Horizons team.

\begin{figure} 
\begin{subfigure}{\linewidth}
\centering
    \includegraphics[width=1.0\linewidth]{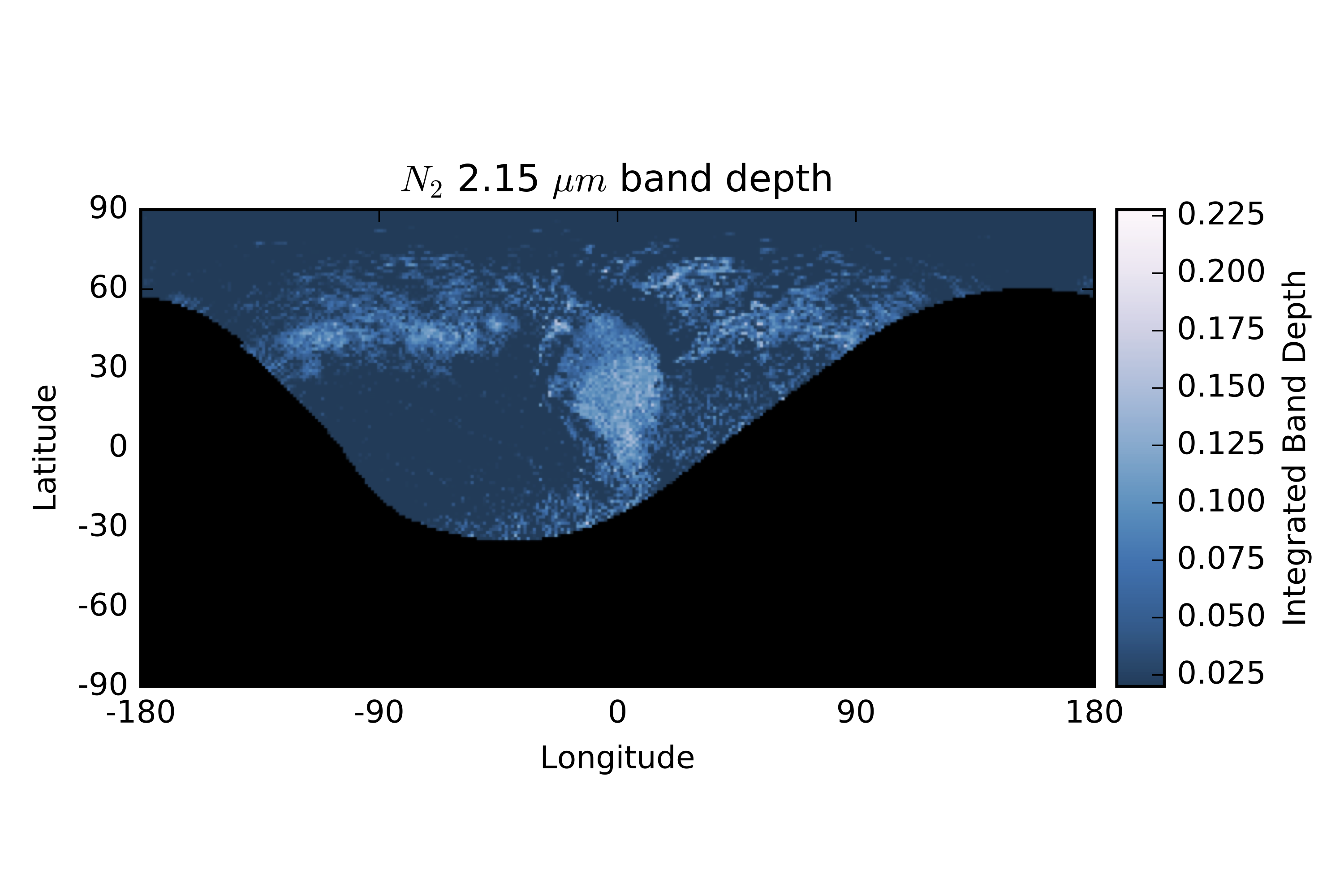}
    \vspace{-80pt}
\end{subfigure}
\begin{subfigure}{\linewidth}
    \centering
    \includegraphics[width=1.0\linewidth]{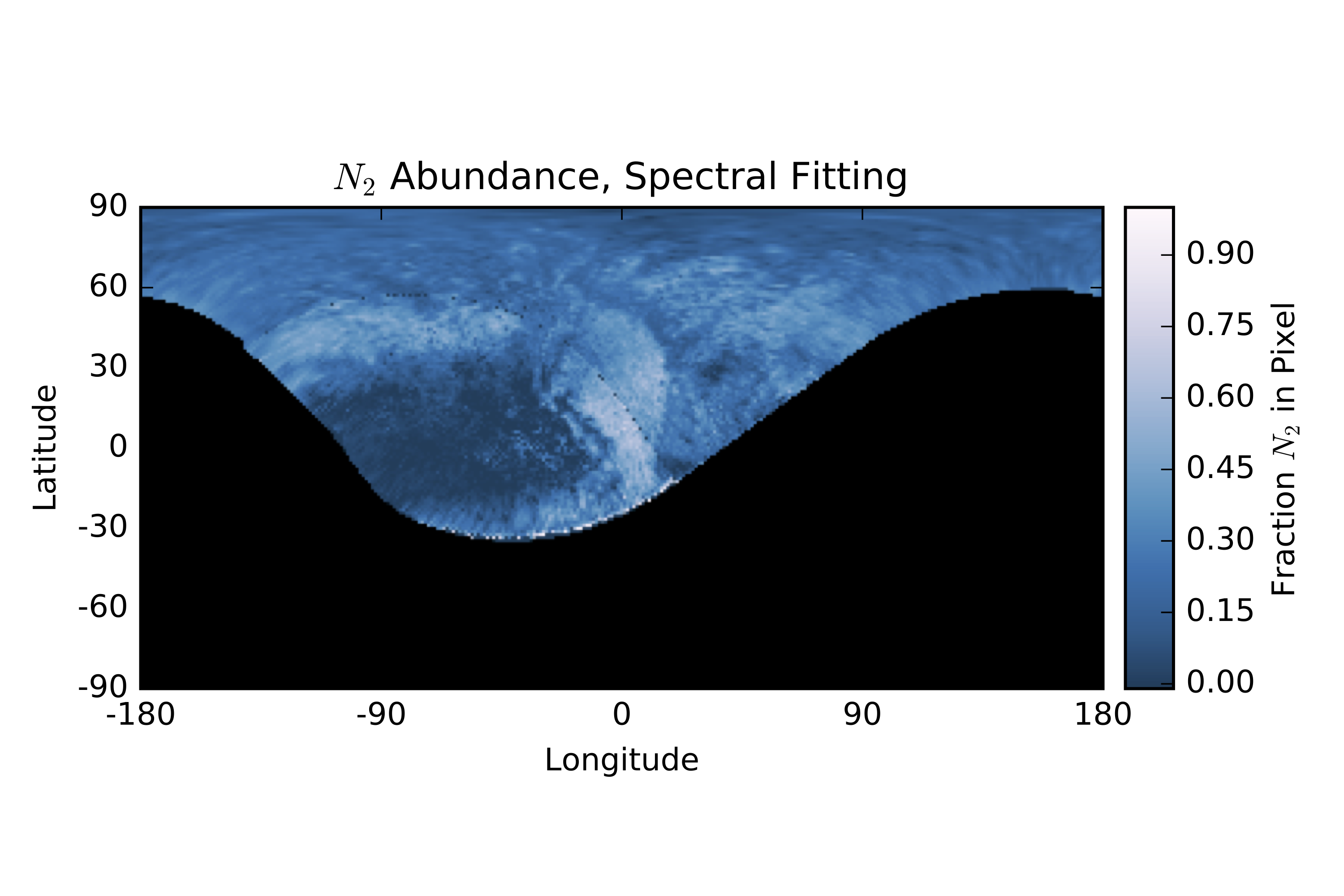}
    \vspace{-40pt}
\end{subfigure}
\caption{Previously published maps of \nit\ ice distribution. Top: Map of the 2.15-\mum\ feature band depth from \citet{schmitt2017physical}. Bottom: Abundance of \nit-enriched component from the spectral model fits of \citet{protopapa2017pluto}. Those fits are strongly influenced by the \nit\ ice 2.15-\mum\ feature, and also include the effects of multiple shifted bands of \met\ dissolved in \nit\ ice.}
\label{fig:prev_comp_maps}
\end{figure}


\section{Geographic and Topographic Distributions of Volatile Ices}\label{methods}

\subsection{Global Composition Maps and the Distribution of \nit\ Ice} \label{sec:comp}

Data from the LEISA spectral imager were previously analyzed to derive detailed spectral maps for Pluto's encounter hemisphere. The LEISA spectra cover wavelengths from 1.25-2.5 ${\mu}$m, with a spectral resolving power $\lambda/{\Delta\lambda}\simeq 240$ \citep{reuter2008ralph}. Multiple absorption features of \met\, CO and H$_2$O ices are present in that wavelength range, as is the overtone band of \nit\ ice at 2.15-\micron. The encounter hemisphere data from LEISA have typical spatial resolution of 6 to 7 km per pixel \citep{grundy2016surface}. There are two main approaches to interpreting those data: directly measuring integrated band depths from the spectra, and fitting spectral models. 

Integrated band depth maps \citep{schmitt2017physical} directly measure the strength of selected absorption features. The approach used in that study enabled removal of some instrumental artifacts (via principle component analysis) not corrected by the data pipeline. The resulting maps retained the full spatial resolution of the LEISA data. Because the weak 2.15-\micron \nit\ band occurs at wavelengths co-incident with the much stronger 2.2-\micron \met\ band, the strength of the 2.15-\micron \nit\ band is quite sensitive to the presence of \met. In areas with a particularly high abundance of \met, it becomes very difficult to discern the \nit\ absorption.
Figure \ref{fig:prev_comp_maps} (top) shows the \nit\ band depth map of \cite{schmitt2017physical}. For our energy balance calculations, we used a new binary map of $N_2$ distribution based on both the observed 2.15-\micron\ band depth and the presence of shifted bands of $CH_4$ diluted in $N_2$ (B. Schmitt, in preparation). This map does not have information on the strength of the \nit\ signatures, but instead provides information on whether or not \nit\ is present in a particular pixel. For the topographic investigation (Section \ref{sec:elev}), we used the band depth only map from \citet{schmitt2017physical}, with a detection threshold of 0.005.


By fitting spectral models to the data, it is possible to simultaneously map \nit\ and \met, even in areas where the \met\ absorption is very strong. \cite{protopapa2017pluto} fit Hapke spectral models to the LEISA encounter-hemisphere data. That approach is computationally expensive, so the models were fit to data that were spatially binned by a factor of 4 (giving maps with a resolution of 24 -- 28 km). These maps can provide more information about the distribution of the spectral components included in the models  (\nit, \met, H$_2$O and tholin), including the fractional abundance and grain size (related to the apparent path length) of each spectral component within each (binned) pixel.

\begin{figure} 
\centering
\begin{subfigure}{\linewidth}
\centering
    \includegraphics[width=0.95\linewidth]{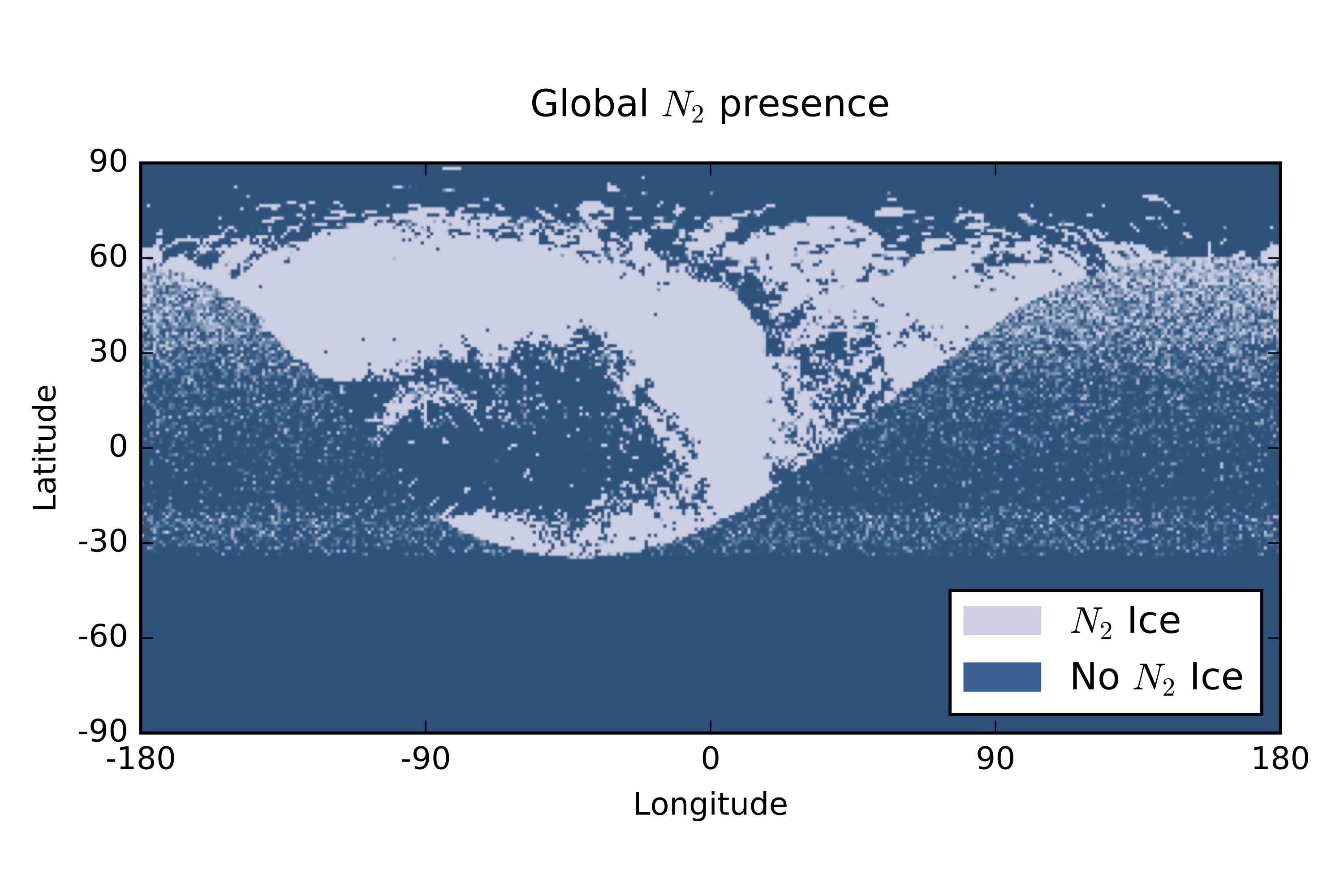}
\vspace{-20pt}
\end{subfigure}
\begin{subfigure}{\linewidth}
    \centering
    \includegraphics[width=0.9\linewidth]{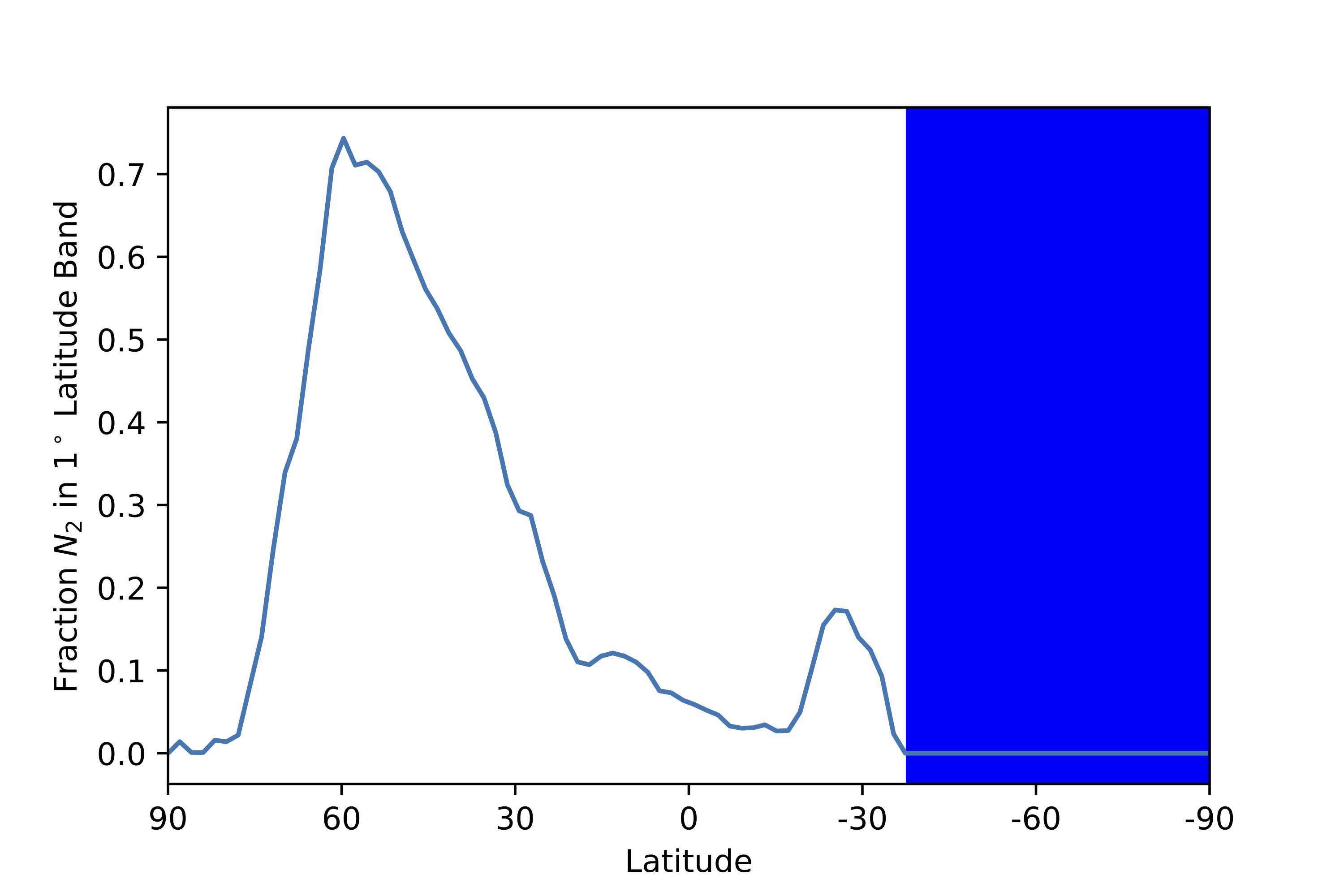}
\end{subfigure}
\caption{(Top) Binary map of nitrogen presence on Pluto based on the results of \cite{protopapa2017pluto} and \cite{schmitt2017physical} (shown in \ref{fig:prev_comp_maps}). The non-encounter hemisphere sections (no-data areas) are filled in by matching the fractional abundance at that latitude, as described in Section \ref{sec:comp}. Areas south of approximately $38^{\circ}$S also have no available data. Here we show that area as being devoid of \nit\ ice, but in Section~\ref{sec:albedo} we explore the implications of different assumptions about \nit\ ice coverage in the un-illuminated southern regions.
(Bottom) Map of latitudinal dependence of Nitrogen ice coverage. Values listed are the fractional abundance of nitrogen ice in each 1 degree latitude bin (e.g. how many pixels of the LEISA data contain the signal of nitrogen out of the total number of pixels in that latitude band). Areas below 38\deg S were unilluminated, so the amount of \nit\ ice there is unknown.}
\label{fig:n2-coverage}
\end{figure}

To determine the albedo and energy balance of just areas with \nit-ice deposits, we required an accurate map of the \nit\ distribution on Pluto's surface. To accomplish this, we dealt with two main challenges: reconciling the band depth maps and the spectral modeling maps, including their spatial-resolution differences, and extrapolating the measured \nit\ distribution to fill in areas beyond the encounter hemisphere and areas on Pluto that were un-illuminated at the time of the encounter.

Firstly, we re-sampled the \citet{protopapa2017pluto} spectral modeling maps to match the resolution of the \citet{schmitt2017physical} integrated band depth maps and qualitative $N_2$ abundance maps. This is equivalent to assuming that the entire pixel in the lower resolution maps of \citep{protopapa2017pluto} contains some of the species (\nit:\met) being mapped. We then combined the maps to create a map of \nit\ ice distribution, as follows. If a pixel in either of the maps (binary observed $N_2$ distribution, $N_2$ abundance from spectral modeling) indicated nitrogen presence, that pixel in our final map was set to 1. For the band depth maps, the acceptable band depth range to indicate nitrogen presence is a value greater than or equal to 0.005. These band depth maps were only used for the investigation in Section \ref{sec:elev}. For all energy balance calculations, we used the binary $N_2$ map that includes $N_2$ detected by both the $CH_4$ band shift and the 2.15-$\mu{m}$ $N_2$ feature. For the spectral modeling maps, any \nit\ abundance above 0.0 indicates presence. Areas without nitrogen present in any of the maps were set to 0. This approach provides a more complete and detailed map of the presence of \nit\ ice on Pluto, but may still underestimate the total inventory because of the weakness of the 2.15-\mum\ band and finite sensitivity of the data. 

To fill in our \nit\ map in areas outside the encounter hemisphere, for which no composition maps have yet been created, and where spatial resolution will be much lower, we adopted a scheme of fractional abundances. On the encounter hemisphere, for each degree of latitude, we counted all pixels covered by nitrogen and divided that by the total number of pixels for which we had data, providing a fraction of the surface that is covered at that latitude, as shown in the bottom panel of Figure \ref{fig:n2-coverage}. We then created a random distribution with the same fraction of pixels covered and inserted that into the no-data areas at that latitude bin to create a global map of nitrogen presence, as shown in the top panel of Figure \ref{fig:n2-coverage}. When calculating the fractional coverage at a given latitude, we excluded pixels falling within Sputnik Planitia. As described below, the process responsible for the \nit\ deposit in S.P. is driven by topography, not seasonal nor climatic effects that should produce latitudinal patterns in the ice distribution. There is no evidence for such large impact basins to be present on the the Charon-facing hemisphere of Pluto \citep{grundy2010tritspex, 2010Natur.468..775S}, so including S.P. in our calculation of the latitudinal distribution of \nit\ would have biased our result. The resulting map of \nit\ ice on Pluto is given in Figure~\ref{fig:n2-coverage}, top panel.


\subsection{Elevation and Volatile Ice Distribution} \label{sec:elev}

Vapor pressure equilibrium has long been understood to govern the atmospheric pressure of Pluto (and similar bodies, Triton certainly, and perhaps Eris, Sedna, and Makemake at their warmer seasons). The concept of vapor pressure equilibrium between global \nit\ ice deposits and Pluto's global-scale atmosphere is complicated by the fact that the \nit\ occurs at a range of elevations (or more precisely, geopotential radii). As pointed out by \citet{stansberry2014plutotopo} and \citet{trafton2015departure}, this produces an additional term that causes the \nit\ ice at higher elevations (lower atmospheric pressure) to sublimate, and for condensation of \nit\ ices to be favored at lower elevations (higher atmospheric pressure), other aspects governing the energy balance of the ice being equal. This basic physical process of downward volatile transport is also confirmed by global climate models of Pluto's surface and atmosphere \citep{bertrand2016atmotopo}. So, Pluto's atmosphere is characterized by both a global vapor pressure equilibrium temperature (\tnit) and equivalent surface pressure (\pnit), and by a global vapor pressure equilibrium radius or elevation (\znit). Any \nit\ ice at elevations other than \znit\ will tend to undergo downward topography-driven volatile transport.

\begin{figure}   
\centering
    \includegraphics[width=1.0\linewidth]{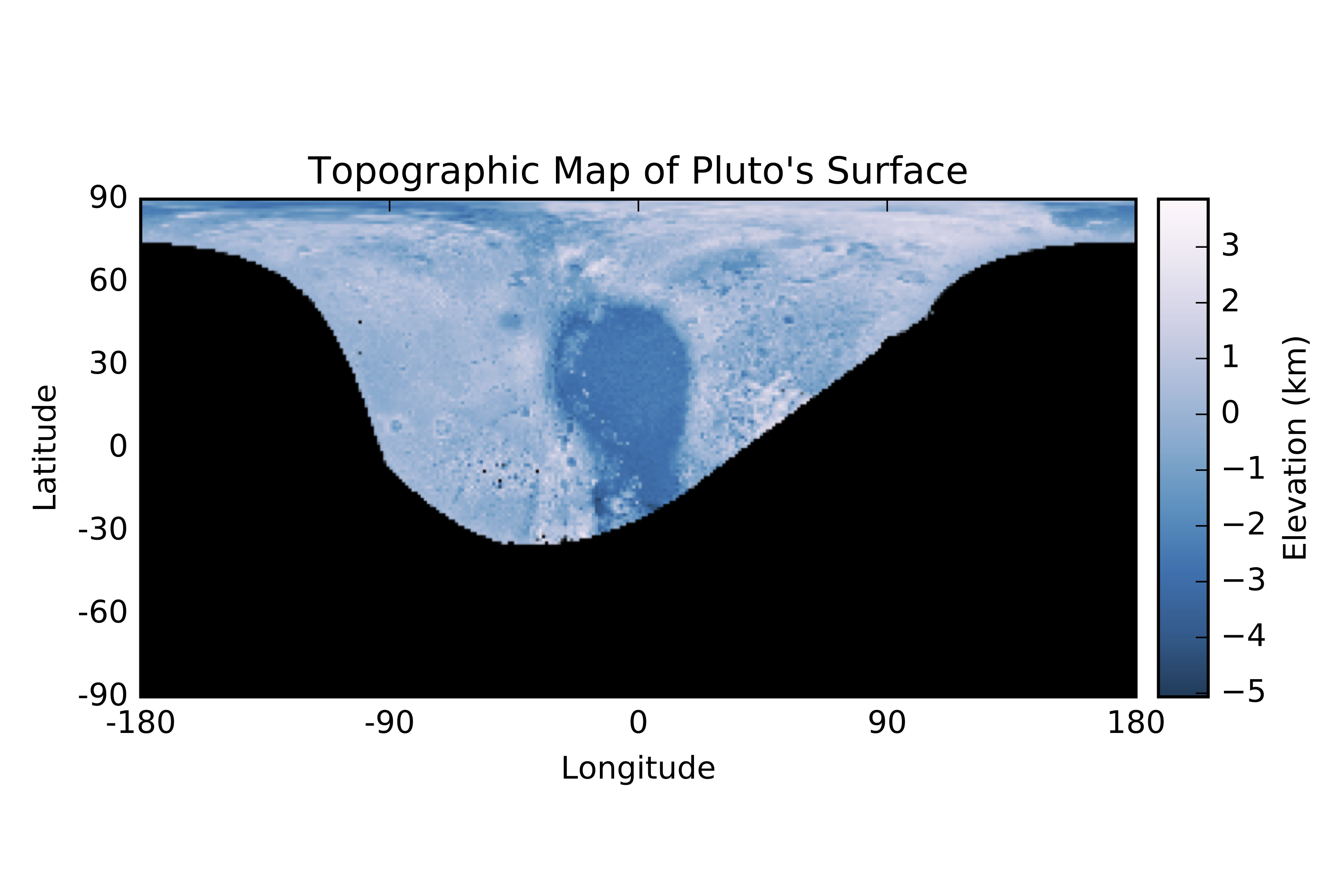}
    \vspace{-20pt}
    \caption{Previously published topographic map of Pluto's encounter hemisphere \citep{schenk2018plutotopo}. The prominent Sputnik Planitia basin near map center dominates the topography. The elevation data are defined such that elevations of 0~km are at the the mean planetary radius.}
\label{fig:z_map}
\end{figure}

The correlation of the huge deposit of \nit, CO and \met\ ice that makes up Sputnik Planitia with low elevations was noted by \citet{grundy2016surface}, and is readily apparent by comparing Figures \ref{fig:prev_comp_maps} and \ref{fig:z_map}. Topography-driven volatile transport of \nit\ on Pluto is relatively slow (for elevation differences of a few km) compared to seasonal transport. This is because topography-driven transport is energy-limited by the difference in thermal emission at \tnit\ and at temperatures given by the wet adiabatic lapse rate at elevations other than \znit. The balance between insolation, emission and latent heat flux in fact drive the surface temperature to the wet adiabat, which is the expression of vapor pressure equilibrium in the presence of a vertical pressure gradient. Thus the timescale for depositing 1 km of \nit\ ice in the Sputnik Planitia basin is of order 1~Ma \citep{trafton2015departure}, while the time to deposit several kilometers is 10 to 30~Ma because the rate of infilling is proportional to the ever-decreasing depth as the basin is filled \citep{bertrand2016atmotopo}. However, once formed, such a large deposit has an important influence on the global energy balance of Pluto's \nit\ ice. On shorter timescales, seasonal transport can locally sublimate or deposit  approximately 1-meter thick layers of \nit\ ice over a Pluto year.

\begin{figure} 
    \centering
    \begin{subfigure}{\linewidth}
    \centering
        \includegraphics[width=0.75\linewidth]{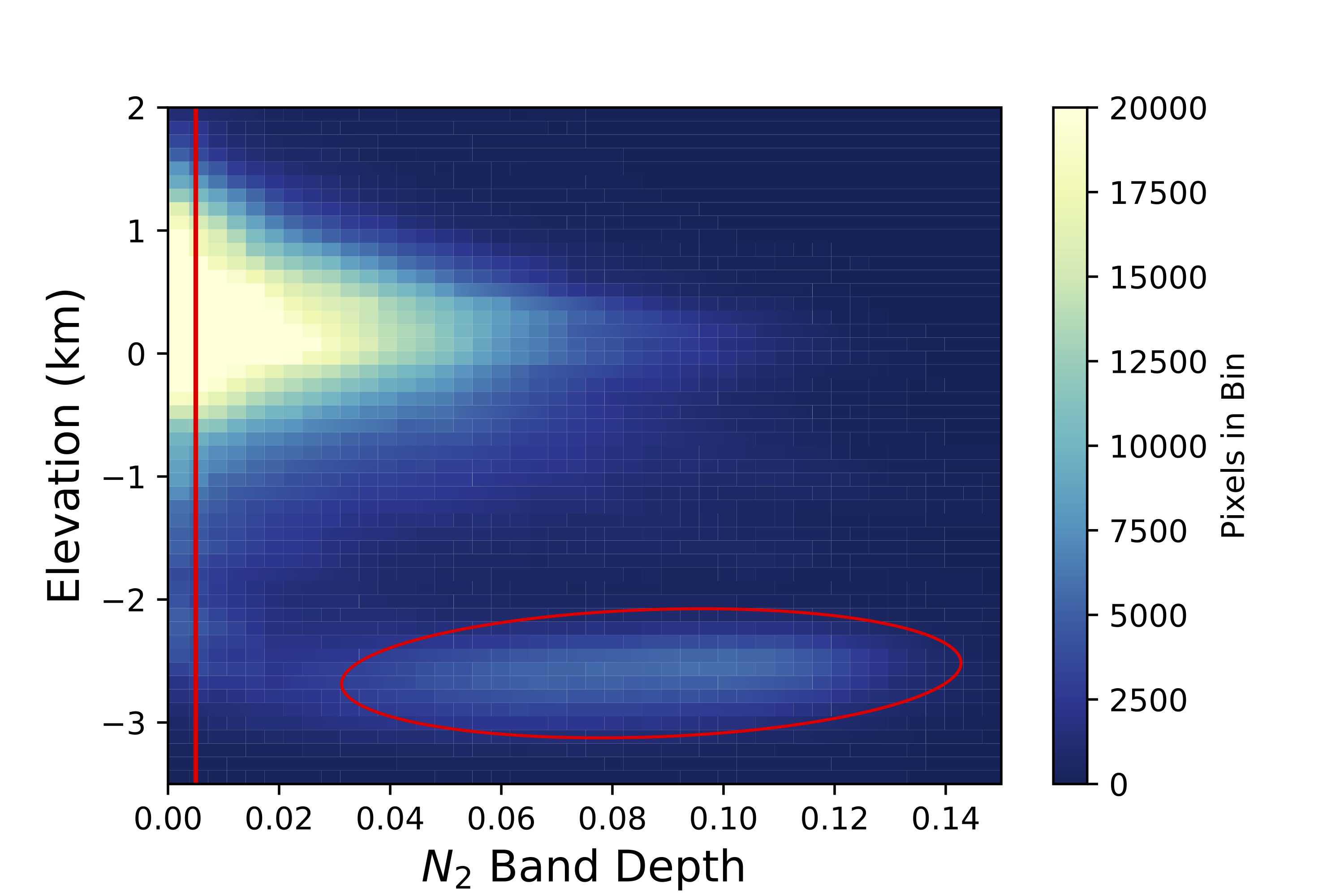}
    \end{subfigure}
    \begin{subfigure}{\linewidth}
    \centering
        \includegraphics[width=0.75\linewidth]{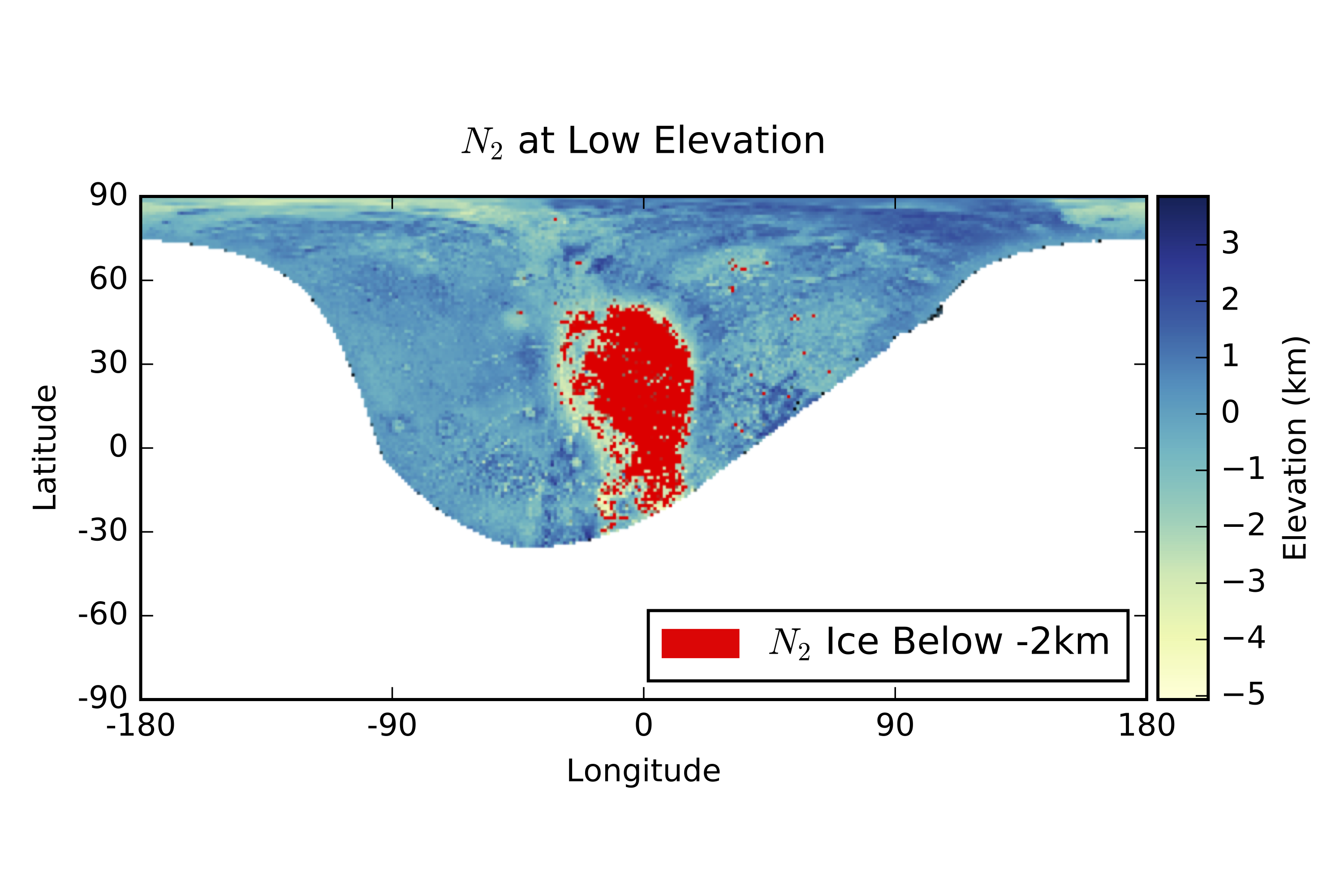}
    \end{subfigure}
    \caption{Histogram of integrated nitrogen band depth \citep{schmitt2017physical} vs. elevation (top). The significant clump of strong \nit\ absorption at elevations near -3~km is mostly due to \nit\ ice in Sputnik Planitia (map of points in this region of parameter space shown on bottom panel, circled on top panel). A second band of \nit\ occurs near 0~km, but does not extend to the strongest absorption strengths seen in Sputnik. Band depth values less than 0.005 (as indicated by the vertical red line, top) are spurious.}
    \label{fig:n2-vs-z}
\end{figure}

The competing influences of seasonal and longer-term topographic volatile transport are apparent in the distribution of Pluto's \nit\ and other volatile ices, as measured by New Horizons. In Figure~\ref{fig:n2-vs-z} we present a histogram of elevation and band depth of the 2.15-\mum\ absorption band of \nit. Here we used only the band depth map to explore the relationship between elevation and \nit\ distribution because the band depth includes both grain size and fractional abundance on the spectrum. Had we used the maps of \citet{protopapa2017pluto}, it would have been necessary to consider both grain size and fractional abundance, resulting in 3-D histograms of the topographic distribution. For the \met\ distribution, we additionally made use of the \citet{protopapa2017pluto} \nit:\met\ dilution ratio map.  
Figure~\ref{fig:n2-vs-z} suggests that there are two main reservoirs of \nit\ ice on Pluto: Sputnik Planitia (at an elevation around -3~km and with the strongest band strengths), and the rest of the planet, with \nit\ occurring at 0~km elevation and in deep craters, indicative of topographic-driven transport. In terms of area, Sputnik Planitia represents a fairly small fraction of the \nit\ ice on Pluto, but the strength of the 2.15-\mum\ band is systematically stronger there than in other regions. The more widely distributed \nit\ at higher elevations has weak absorption strengths, and is predominantly in areas with high \met\ abundance (as indicated by the \nit\ band strength being near zero, see Figure 18 of \citet{schmitt2017physical}).

\begin{figure} 
    \centering
    \begin{subfigure}{\linewidth}
        \centering
        \includegraphics[width=0.75\linewidth]{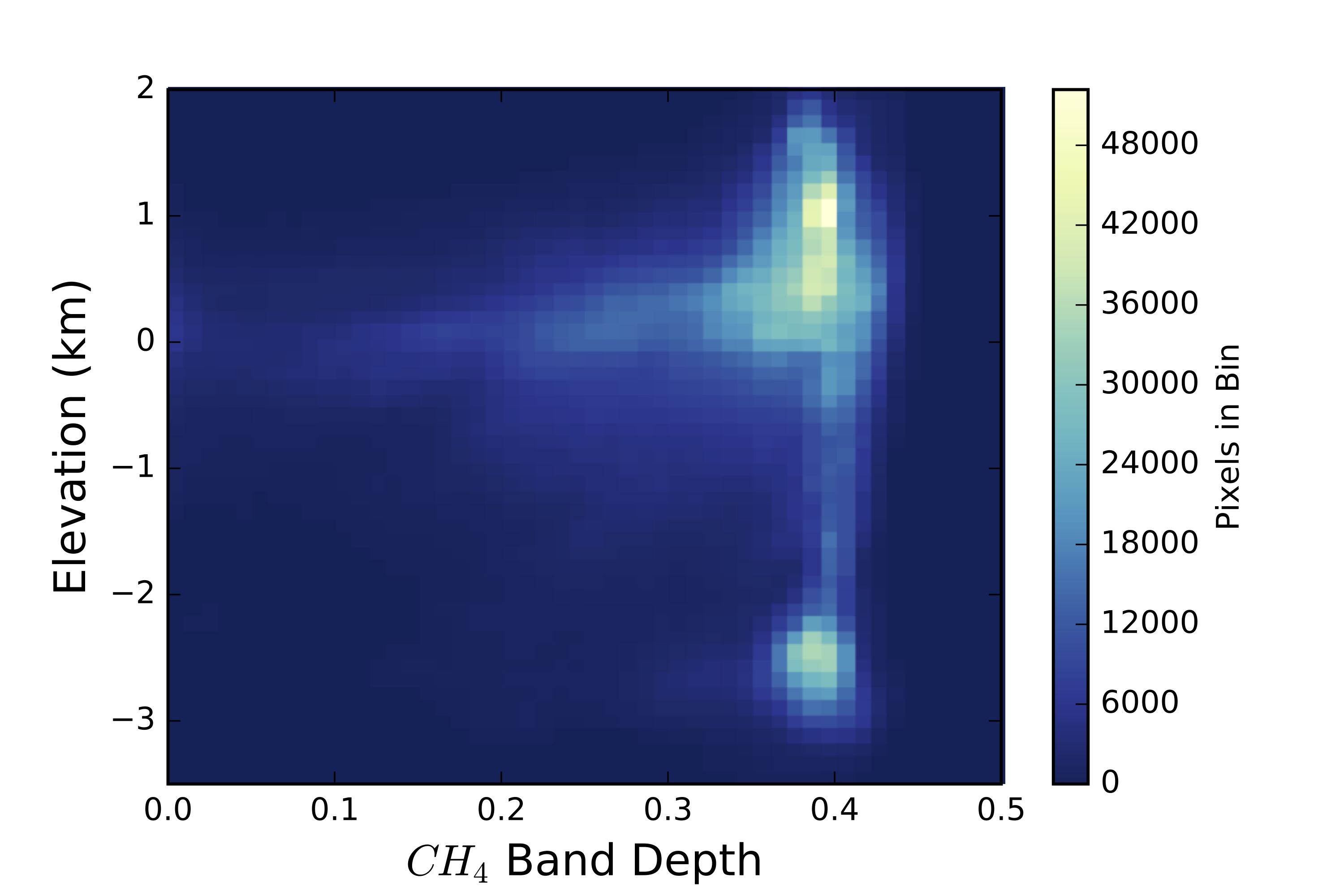}
    \end{subfigure}
    
    \begin{subfigure}{\linewidth}
    \centering
        \includegraphics[width=0.75\linewidth]{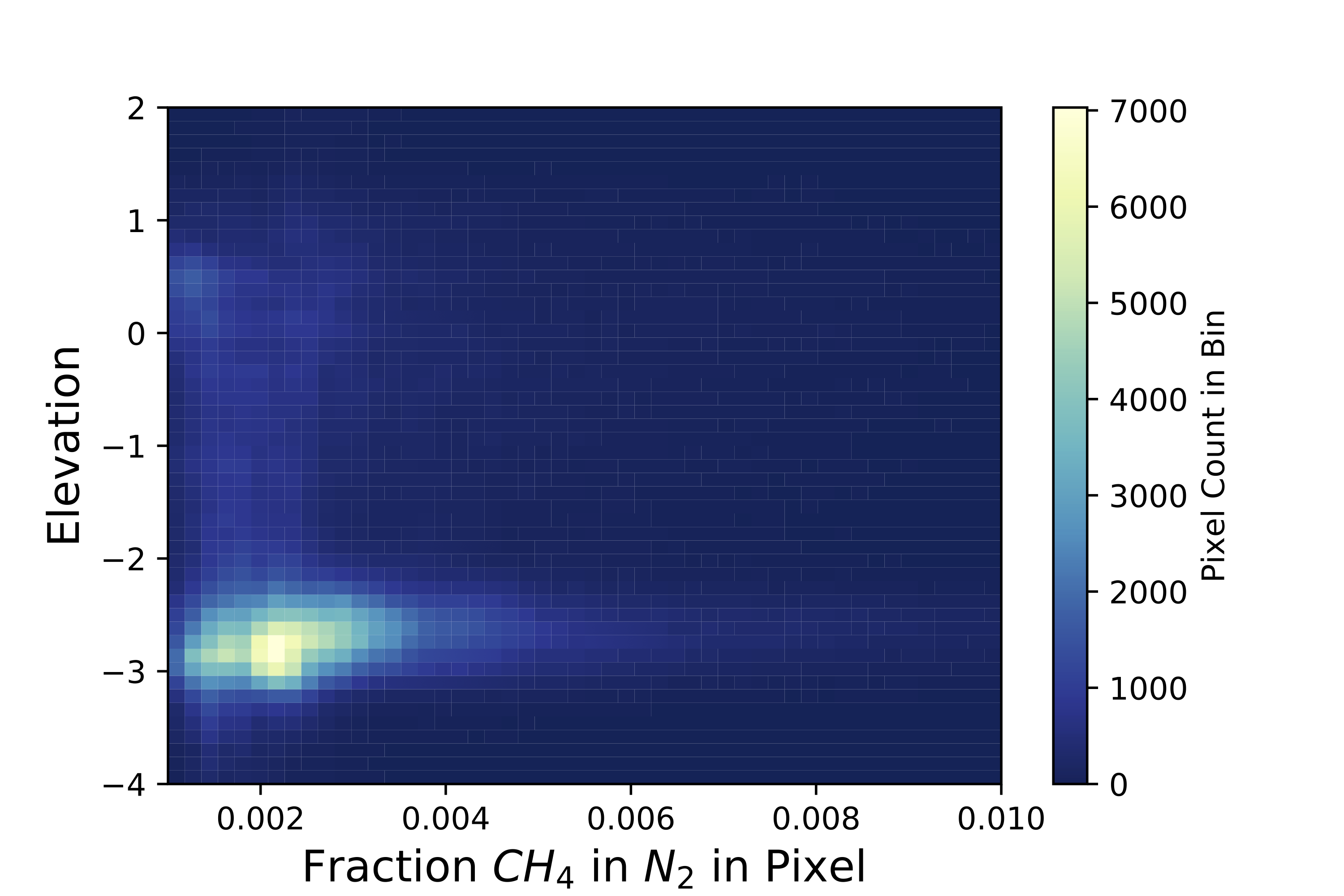}
    \end{subfigure}
    
    \vspace{-0pt}
    \caption{Top: 2-D histogram of integrated methane band depths \citep{schmitt2017physical} vs. elevation. The clump of pixels at -3\,km elevation maps to Sputnik Planitia. Bottom: 2-D histogram of \met\ diluted in \nit\ vs. elevation \citep{protopapa2017pluto}. Methane present in solution with nitrogen shows clear evidence of downward topographic transport.  
    }
    \label{fig:ch4-vs-z}
\end{figure}

Methane also participates in volatile transport on Pluto, albeit not as a primary constituent of the atmosphere \citep{hinson2017radio,young2017n2}. Figure~\ref{fig:ch4-vs-z} is the 2-D histogram of elevation vs. \met\ band depth. Here we rely on the \citet{schmitt2017physical} \met\ maps shown in their Figure~13. 
Unlike the \nit\ band depths shown in Figure~\ref{fig:n2-coverage}, the detections of \met\ are robust across Pluto's surface. The distribution reveals something of a dichotomy, as does that of \nit. Sputnik Planitia produces a significant cluster of pixels with strong absorption at an elevation of -3~km, and a second, larger cluster with (in some cases) weaker absorption occurring near 0~km elevation. However, \met\ also occurs at higher elevations than \nit\ directly detected via the 2.15-\mum\ absorption feature, and at higher elevations (up to nearly 2~km) the bands are just as strong as those in Sputnik. As shown in the lower panel of Figure \ref{fig:ch4-vs-z}, \met\ diluted in \nit\ is mostly concentrated at low elevations. 

\begin{figure} 
    \centering
    \includegraphics[width=0.75\linewidth]{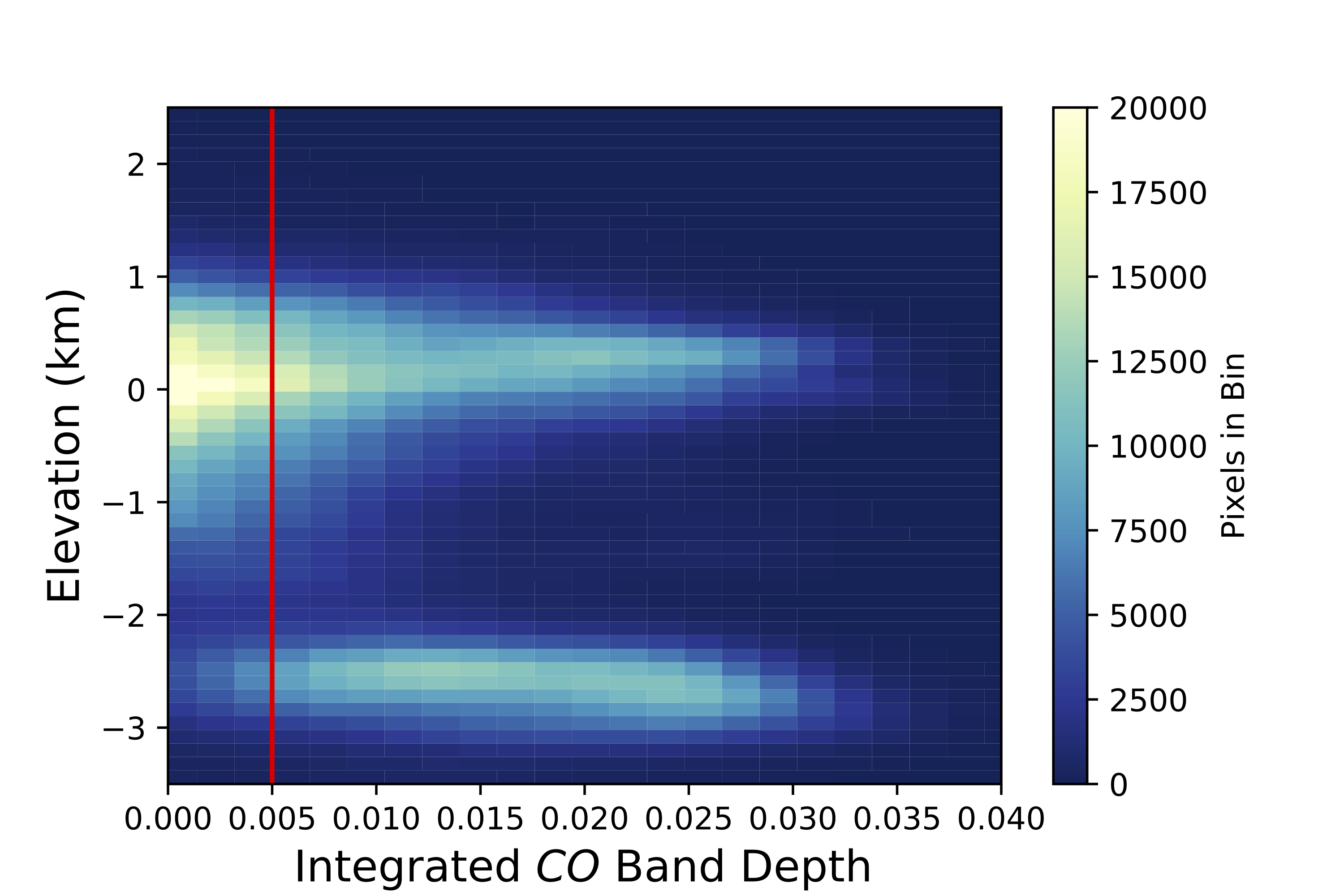}
    \caption{Histogram of integrated carbon monoxide band depth \citep{schmitt2017physical} vs. elevation. The cluster of pixels at -3\,km elevation maps to Sputnik Planitia, similar to the \nit\ distribution shown in Figure~\ref{fig:n2-vs-z}. The vertical red line denotes the limit for reliable, physical band depth measurements (0.005). 
    }
    \label{fig:co-vs-z}
\end{figure}

Figure \ref{fig:co-vs-z} shows the elevation distribution of CO ice, based on the \citet{schmitt2017physical} band depth measurements from very high spectral resolution measurements. Merlin \etal\ (2018) conclude that the width of the CO absorption bands on Triton imply that the CO there is present {\em only} dissolved in \nit\ ice. If that were true on Pluto, the CO absorption band maps could be used as a tracer for \nit, just as the binary \nit\ map used for this study included the band-shifted \met\ absorptions as a tracer of \nit. Indeed, the CO ice distribution in Figure~\ref{fig:co-vs-z} closely resembles the \nit\ ice distribution in Figure~\ref{fig:n2-vs-z}, suggesting a close physical association between these molecules on broad spatial scales. This is also a conclusion of \cite{schmitt2017physical}, which also identified a trend of \nit\ with weak or no CO, corresponding to N. Vega Terra, Hayabusa Terra, and Pioneer Terra. Unfortunately measuring or fitting the CO ice band in the LEISA data is difficult. The band depth data lie primarily below the threshold of 0.005, although valid band depths are seen at both -3\,km (which we have verified to map to Sputnik Planitia) and near 0\,km elevation. Attempts to fit spectral models to that band by \citet{protopapa2017pluto} were frustrated by the lack of adequate optical constants in the wings of the band, and no results for CO were presented there. Thus, while the similarities between the elevation distributions of CO and \nit\ on Pluto are suggestive, CO ice is even harder to detect in the LEISA spectral band than is \nit\ ice making it impractical to use as an additional ``tracer'' for \nit.

In summary, the figures above show that large-scale topography on Pluto does indeed influence the distribution of \nit, \met\ and CO. Nitrogen ice in Sputnik Planitia basin, with both \met\ and CO dissolved in it, is evidence of the slow but steady down-hill transport of those ices caused by the atmospheric pressure gradient, and is not expected to be significantly disturbed by seasonal redistribution of the ices (\citet{bertrand2016atmotopo}). (Below we discuss the distribution of sublimation and condensation of \nit\ within Sputnik Planitia at the time of the New Horizons encounter.) Outside Sputnik Planitia, at least on the encounter hemisphere, the \nit\ ice is distributed approximately in latitude bands that are consistent with expected seasonal transport (\eg\ \citet{young2013vt3d}), with albedo and topography (\eg\ altitude) of the surface modifying the distribution within latitude bands. In other words, lower albedo regions in a given latitude band tend to have less \nit\ than the higher albedo regions. We explore further the albedo of the \nit\ ice in the following section.

\subsection{Total Nitrogen Ice Inventory}

Using these maps and observations, it is possible to put a lower limit on the total amount of \nit\ on Pluto. By multiplying the fractional abundance, pixel area, and equating the grain sizes from \citet{protopapa2017pluto} to the local depth of the \nit\ ice (a severe lower limit on the layer depth due to the transparency of the \nit), and integrating over the map, we derive a lower estimate of the volume of \nit. Grain size values that correspond to a \nit\ abundance less than $10\%$ are invalid, so these were excluded from the calculation (as stated in \citet{protopapa2017pluto}). For reference, typical \nit-ice grain sizes were found to be $\simeq 50$\, cm by \citet{protopapa2017pluto}. Note that this calculation did not include the extrapolated fractional coverage beyond the encounter hemisphere, since we were interested in a lower limit based only on the available data. Through this method, we found that there is approximately $6*10^2$ km$^3$ of \nit-ice distributed across the globe outside of Sputnik Planitia. The total amount of \nit\ on Pluto has implications for understanding how \nit\ was incorporated during the formation of Pluto, as investigated in \citet{lisse2019}.

Is SP the dominant reservoir of \nit, or a relatively equal contributor compared to the rest of the surface? Given that the \nit\ ice layer in Sputnik Planitia is estimated to be kilometers thick \citep{mckinnon2017origin}, so the grain sizes are not appropriate for estimating the \nit\ inventory there. Instead we integrated over the area of the Sputnik basin, assuming a uniform composition of pure \nit\ and a depth of 3 km. This resulted in approximately $3.5\times10^5$ km$^3$ of \nit\ in SP alone. That is 500 times more than the lower bound calculated above for the \nit-ice inventory outside SP Given that the \nit\ ice outside SP generally does not tend to mute, let alone obscure, underlying topography, it seems unlikely that its typical depth would be 500 times greater than assumed above (\ie 250\,m on average). Thus SP likely represents the largest reservoir by volume of \nit\ ice on Pluto by far. 
As such, it isn't surprising that we see evidence of SP playing a key role Pluto's atmospheric cycles \citep{bertrand2018nitrogen}.


\section{Energy balance and transport of Pluto's \nit\ ice}\label{energy}

Using the map of \nit\ ice distribution described in Section \ref{methods}, we examined the global energy balance of the ice. Because the atmosphere is supported via vapor pressure equilibrium with the global \nit\ ice inventory, its energy balance is a key consideration for understanding the overall surface-atmosphere system, and provides insights into processes governing the redistribution of \nit\ (and other volatile ices) on seasonal and climatic timescales. As mentioned in Section \ref{intro} (and based on the extensive discussion in \citet{trafton2015departure}), the expected non-ideal vapor pressure behavior of \nit:\met\ ice might have significant implications for both latent heat redistribution and the link between the temperature of the \nit\ ice and the atmospheric pressure. Lacking any laboratory constraints on those non-ideal effects, we assume for now that the solid solution ice \nit:\met\ ($N_2$-dominant mixture) behaves like pure \nit\ ice.


\begin{figure} 
    \centering
    \includegraphics[width=0.95\linewidth]{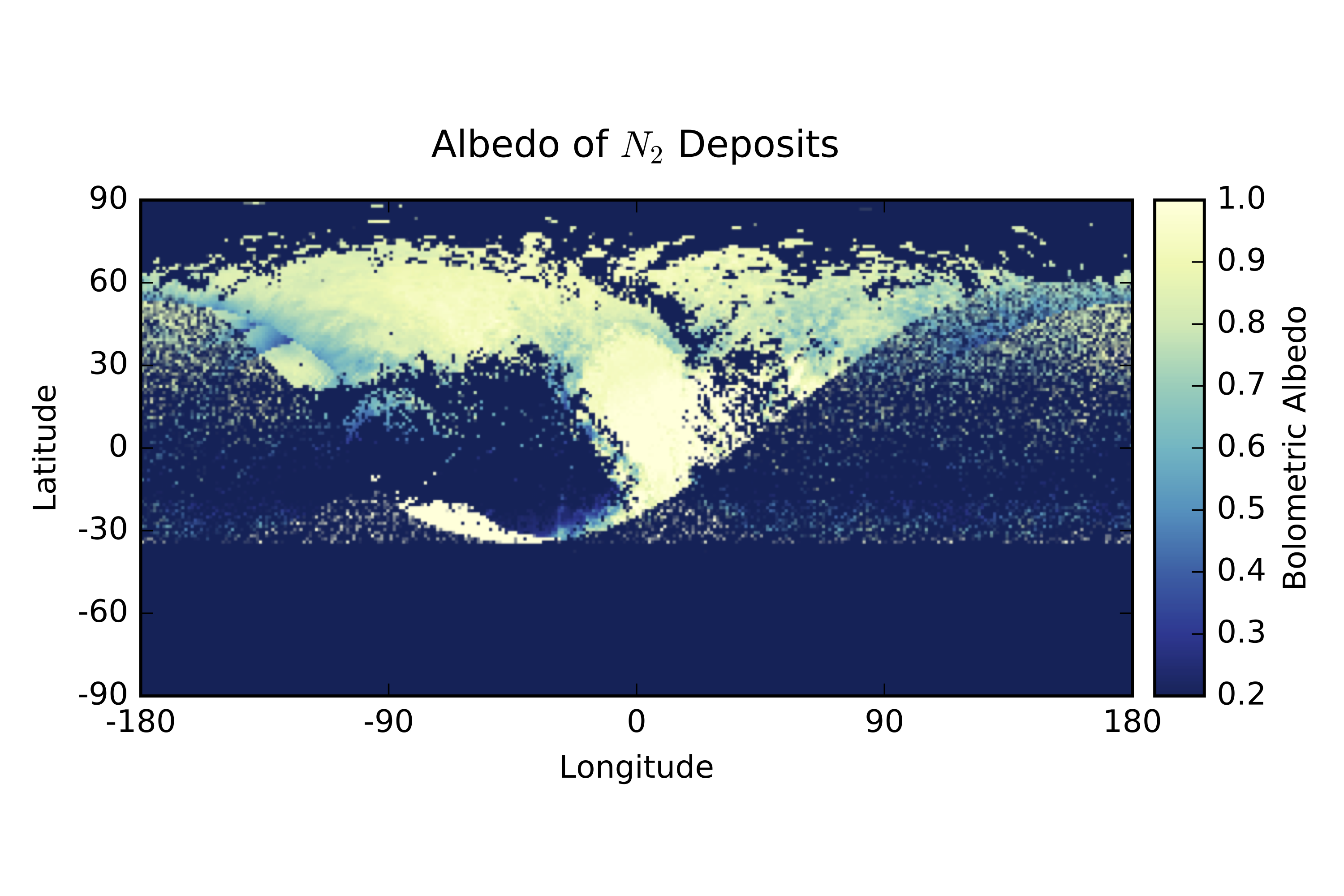}
    \caption{Map of bolometric albedo \citep{buratti2017global} of nitrogen deposits across the surface.  Areas lacking LORRI data, and therefore albedo determinations, but north of -38\deg\ latitude are filled in in the same manner as for the \nit\ presence map of \ref{fig:n2-coverage}, using the average albedo of \nit\ in all other areas.}
    \label{fig:albedo-map}
\end{figure}

\subsection{Bolometric albedo, temperature, and emissivity of the \nit ice} \label{sec:albedo}

Maps of bolometric albedo, determined from LORRI data, over a bandpass of $0.61 \pm 0.2$\mum\ \citep{buratti2017global}, were combined with the global map of nitrogen presence from Section \ref{sec:comp} to create a map of the albedo of nitrogen deposits across Pluto's surface, as shown in Figure~\ref{fig:albedo-map}. These albedos are somewhat approximate, as described in \citet{buratti2017global}. A full analysis of the spatially-dependent phase function of Pluto has not yet been done, so the bolometric albedos shown here are based on the Triton phase function \citep{hillier1990trit}. While this approach may result in some inaccuracies, \citet{buratti2017global} show that Pluto's and Triton's phase functions at the global scale are very similar. Thus, the resulting errors in our calculations of absorbed insolation and energy balance of the \nit\ ice are small in the globally-averaged sense. \citet{hillier1994trit} examined the photometric properties of various terrains on Triton, finding that the bolometric albedos for those terrains likely to be associated with \nit\ ice differed by no more than about 10\%\ from the global bolometric albedos in their 1990 study.

Pluto's \nit\ ice will also absorb insolation at near-IR wavelengths, due primarily to the few percent of \met\ dissolved in it. We estimated the magnitude of this effect using the global-average spectrum of Pluto from 0.7 -- 5.0\mum. We assumed that the \nit\ ice is spectrally neutral below 0.7\mum, and normalized the entire spectrum to an albedo of 0.62 at 0.7\mum. The integral of the product of that spectrum and the solar spectrum was 1.09 times the integral assuming a constant albedo of 0.62 over the entire range. Because it is the dissolved \met\ that must be responsible for the bulk of that additional absorption in this calculation, and the \met\ band strengths in the \nit:\met\ solid solution may be weaker than for the global-average spectrum (because \nit, and not \met, is the dominant species in those ices), the estimate above may be an upper limit on the additional insolation absorbed by Pluto's \nit\ ice in the near-IR.

With the bolometric albedo map, we can now compute the globally averaged energy balance of the \nit\ ice. As discussed earlier, latent heat transport through the atmosphere maintains all of the \nit\ ice at a single temperature (\tnit), and the heat content of Pluto's thin atmosphere is negligible, so the globally averaged energy balance is the prime consideration for computing \tnit. Equation \ref{global-balance} equates the total thermal radiation from the \nit\ ice to the total insolation absorbed by the ice. We assumed that the emissivity was the same for all the \nit\ ice (with an assumed value of unity), and used the solar flux at Pluto's distance ($S_0$) appropriate for July 2015 (the time of flyby) from the JPL HORIZONS ephemeris database. The bolometric albedo of the ice ($A_{N_2}$), was taken from Figure~\ref{fig:albedo-map}.  

\begin{equation}
    \epsilon_{N_2}\sigma{T_{N_2}^4}\int_{N_2}dA = S_0\int_{N_2}(1-A_{N_2}) <{\cos\mu}>\, dA
\label{global-balance}
\end{equation}

The integrals were carried out over the entire surface (left-hand side), and over the lit fraction of the surface (right-hand side), where \nit\ ice was present (Figure~\ref{fig:n2-coverage}). A further simplification was possible because the time scale to change Pluto's atmospheric mass is much longer than Pluto's day, so we used the diurnal average ($<{\cos\mu}>$) of the insolation angle at each latitude. The terminator longitude at each latitude is given by Equation~\ref{betas}, where $\beta_T$ is the angle from the sub-solar meridian to the terminator, $\theta$ is the co-latitude and $\theta_S$ is the co-latitude of the subsolar point. 

\begin{equation}
    \cos{\beta_T} = \frac{-1}{\tan\theta\tan\theta_s}
\label{betas}
\end{equation}

\begin{figure} 
\centering
\begin{subfigure}{\linewidth}
\centering
    \includegraphics[width=0.85\linewidth]{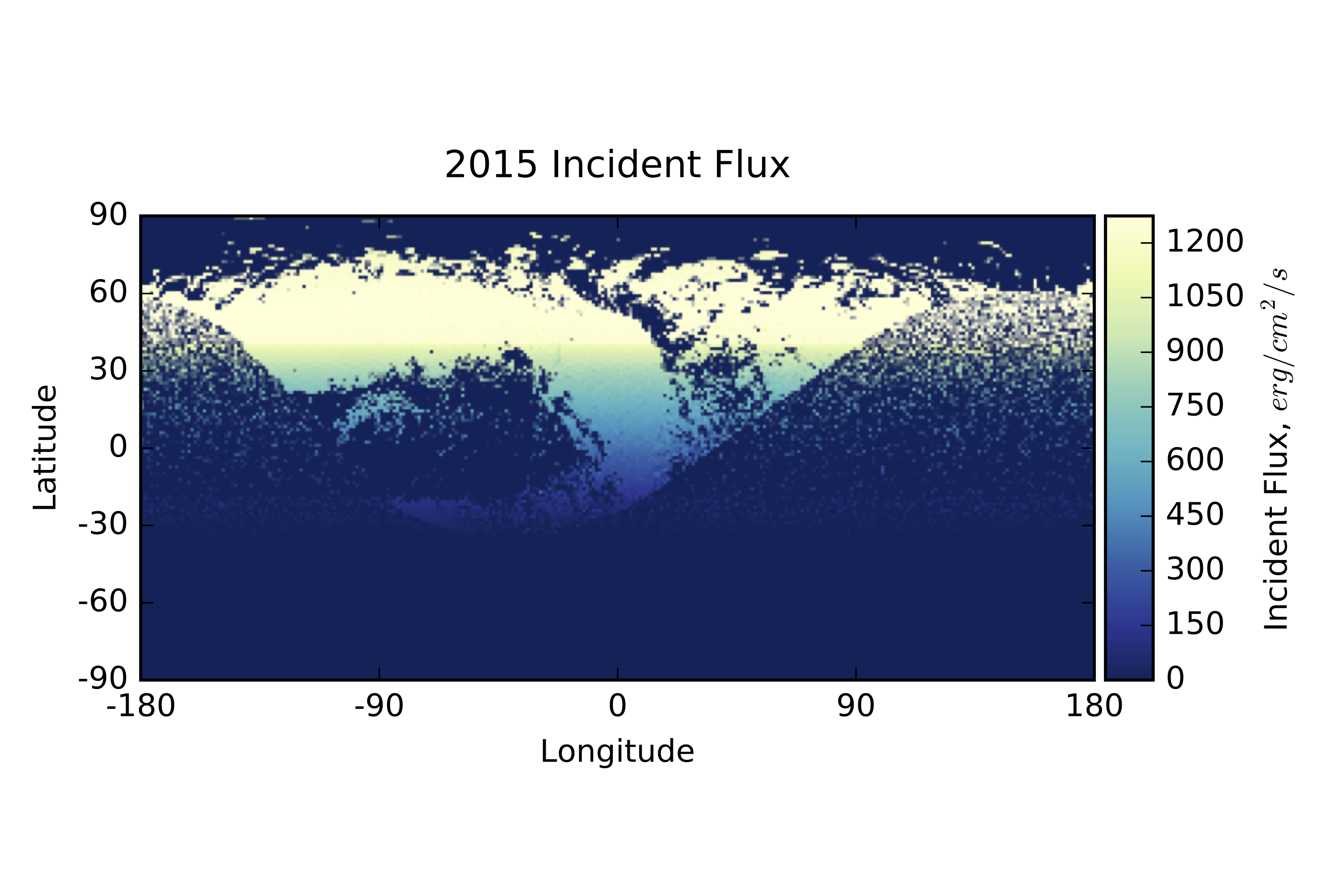}
    \vspace{-40pt}
\end{subfigure}
\begin{subfigure}{\linewidth}
    \centering
    \includegraphics[width=0.85\linewidth]{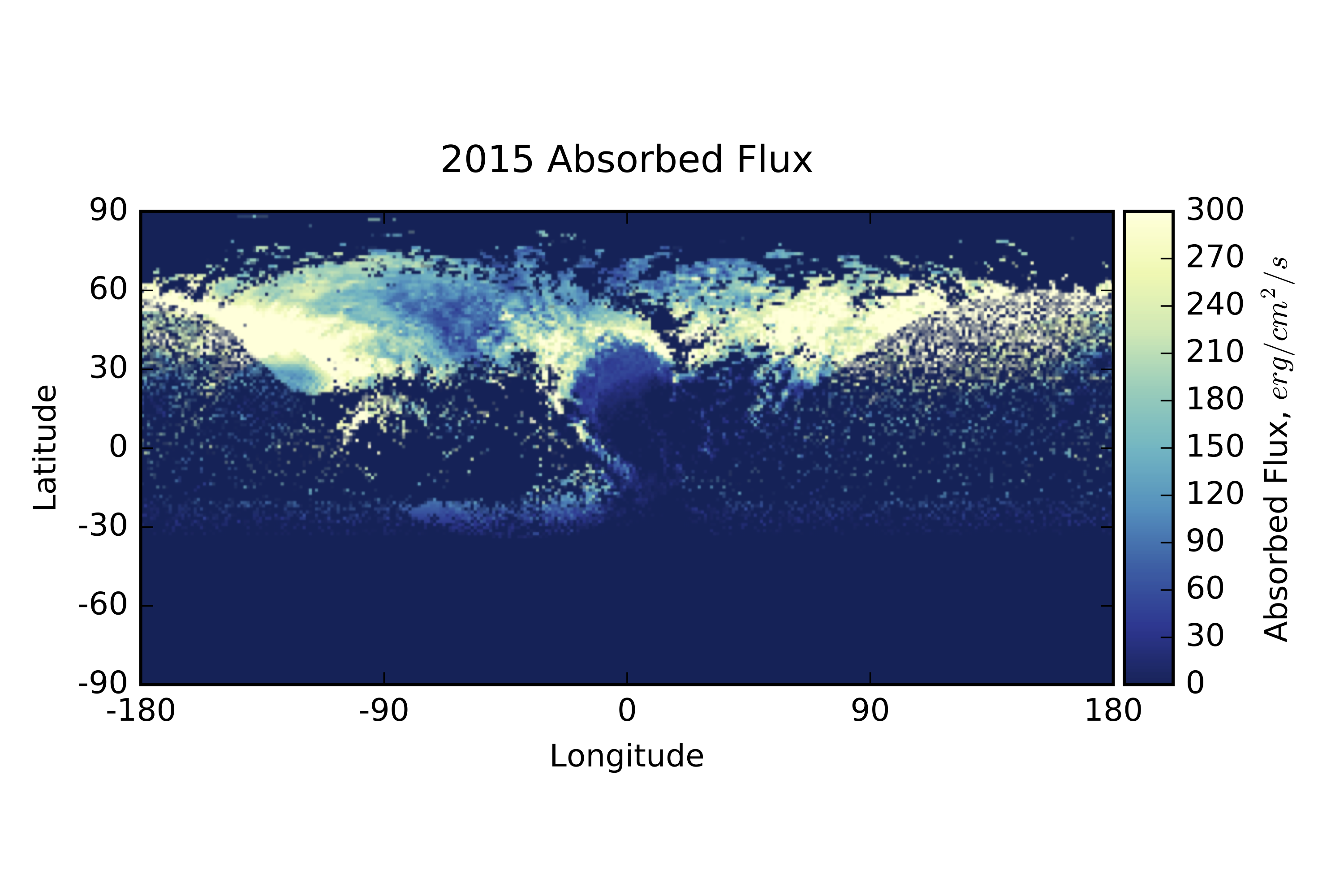}
    \vspace{-30pt}
\end{subfigure}
\caption{(Top) Diurnally-averaged incident solar flux ($< S_0 >$) at the time of the New Horizons flyby, for areas we have identified as having $N_2$ ice. 
(Bottom) Absorbed solar flux at the time of the flyby, as given by $<{S_0}>\, (1-A)$, where A is the bolometric albedo (Figure~\ref{fig:albedo-map}). }
\label{fig:flux}
\end{figure}

Figure~\ref{fig:flux} shows the pattern of diurnally averaged incident and absorbed solar flux for Pluto's \nit\ ice. The flux incident on the \nit\ ice is dominated by  the subsolar latitude (51.6\deg N) at the time of the encounter and the overall presence (and absence) of \nit\ ice. The absorbed insolation is much smaller, illustrating the strong modification of the insolation by the high albedo of the \nit\ ice. 

We solved Equation~\ref{global-balance} for \tnit, and present the implications of the result below. While there are no data constraining the presence of \nit\ ice in the {\em terra incognita} (\ie\ un-illuminated portions of Pluto, below 38\deg S), \nit\ ice present in that area significantly affected our results for \tnit. In the absence of observations of the {\em terra incognita} region, we adopted 5 possible distributions for the \nit\ ice there: 1) no \nit, 2) minimum longitudinal-average coverage, 3) mean longitudinal-average coverage, 4) maximum longitudinal-average coverage, and 5) completely covered in \nit\ ice. (See Figure~\ref{fig:n2-coverage} for the longitudinal average values, and Figure~\ref{fig:mass-flux} for realizations of cases 2 through 4.) The details of the distribution of the \nit\ ice within the un-illuminated areas is unimportant for the global energy balance: only its contribution to the area on the left-hand side of Equation~\ref{global-balance} matters. 

Models of long-term seasonal transport on Pluto combined with the history of atmospheric pressure established by stellar occultations can predict the distribution of the \nit\ ice in the {\em terra incognita}, but are subject to significant uncertainties in model parameters and assumptions (in particular thermal inertia and total \nit\ ice inventory). Model results presented by \citet{young2013vt3d}, \citet{bertrand2016atmotopo} and \citet{bertrand2018nitrogen} tend to favor fairly high thermal inertia (800 -- 3000 \tiu, about 0.3 to 1 times that of solid \wat-ice) because they better match the sustained increase in Pluto's atmospheric pressure observed between 1989 and now. Those model cases predict little or no \nit\ ice in the far south at the time of the New Horizons encounter. However, some model runs in \citet{young2013vt3d} (particularly a few of the ``exchange with pressure plateau'' runs) result in some \nit\ ice south of -38\deg. \citet{forget2017post} and \citet{bertrand2017haze} present GCM models for Pluto's atmosphere in 2015 and find that the presence of some \nit\ ice in the far south may be more consistent with the inferred meridional circulation of Pluto's atmosphere. Thus, while seasonal transport modeling can, subject to uncertainties in several model parameters, predict the distribution of \nit\ ice in the {\em terra incognita}, at this time the range of model predictions does not strongly constrain it. We note that the unknown altitude of terrains south of -38\deg\ could have a controlling influence on the distribution of \nit\ ice there: a large basin would, like SP, be filled with \nit\ while if that region is systematically higher than \znit\ it would be less prone to condensation of \nit. The cases outlined above span this range of possibilities, but in the absence of any actual constraints on the overall topography in the {\em terra incognita} we don't assume a strong influence by topography there.

\begin{figure} 
\centering
    \includegraphics[width=0.9\linewidth]{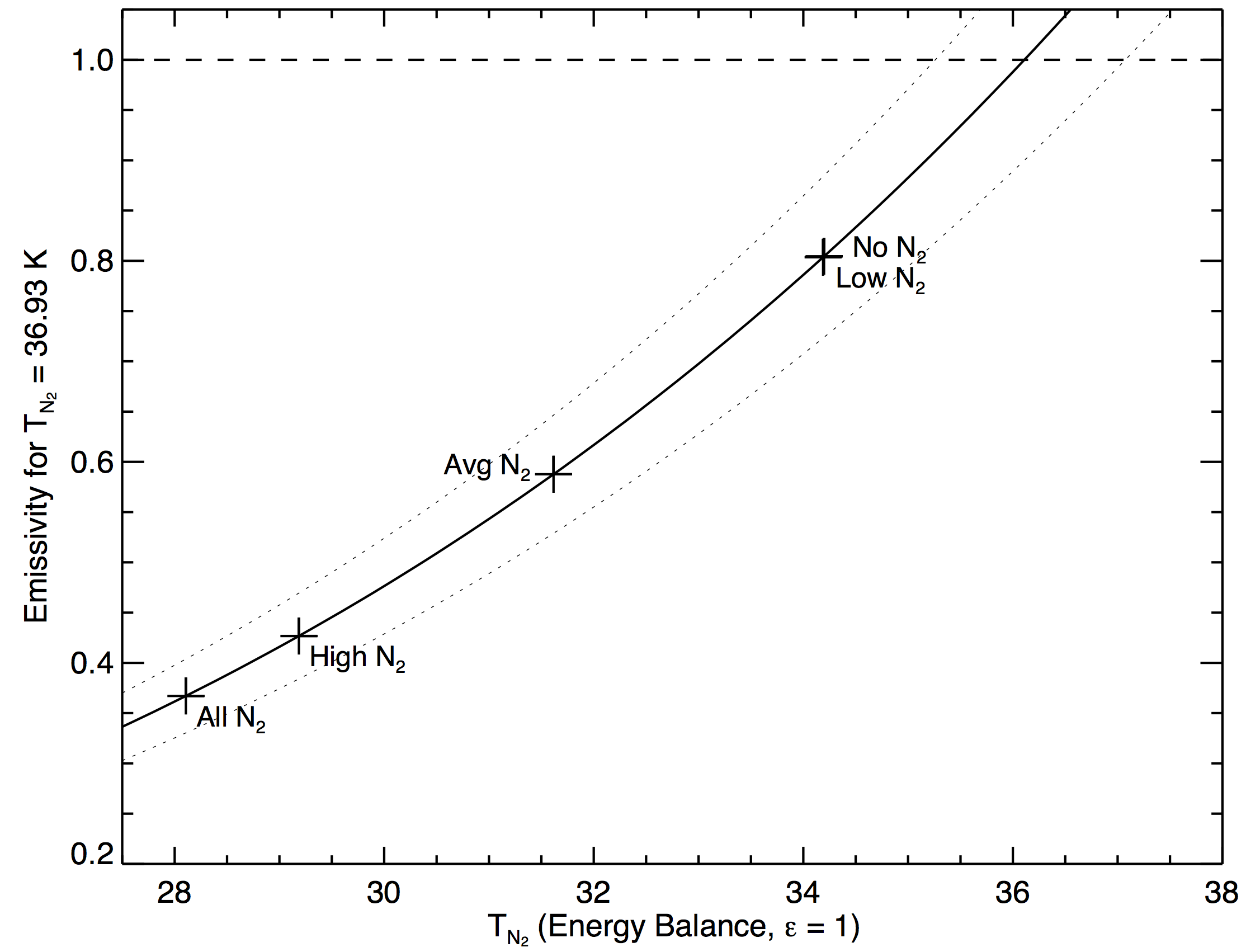} 
    \vspace{-10pt}
\caption{Bolometric emissivity of Pluto's \nit\ ice required to match the temperature (\tnit$ = 36.93K$) inferred from the atmospheric pressure of 11.5\mub\ (solid line). Results of our energy balance calculations (which assumed emissivity = 1) are overplotted as crosses, with labels indicating the 5 cases of assumed distribution of \nit\ ice in the un-illuminated regions south of -38\deg\ latitude. Dotted lines indicate an assumed systematic uncertainty of 10\% on the emissivity we derive, while the horizontal dashed line at an emissivity of unity marks the physical upper limit.}
\label{fig:emissivity}
\end{figure}

Figure~\ref{fig:emissivity} shows the results of our energy balance calculations for Pluto's \nit\ ice. The 5 assumed configurations of \nit\ ice in the un-illuminated regions in the south are plotted as symbols, with the value of \tnit\ for each configuration on the x-axis. For each model, we calculated the bolometric emissivity of the \nit\ ice required to give \tnit$ = 36.93$\,K, (the emissivity was assumed to be the same for all of the \nit\ ice). These results, like the GCM and seasonal models summarized above, suggest that significant deposits of \nit\ ice in the southern {\em terra incognita} region are unlikely. For cases 5 (`all-\nit') and 4 (`high-\nit'), the bolometric emissivity would be 0.32 to 0.46, straining the bounds of plausible emissivities. For our case 3 (`average \nit'), the range is 0.53 to 0.65, while for cases 2 and 1 (`low-\nit' and `no-\nit'), the range is 0.70 to 0.86. These ranges include an assumed 10\% assumed systematic uncertainty on the emissivity, consistent with likely uncertainties in the bolometric albedos and the accuracy of our \nit\ ice occurrence map. Based on the GCM and volatile transport modeles discussed above, it seems most probable that there is some \nit\ ice in the {\em terra incognita}, perhaps consistent with our case 3. If we include the upper emissivity bound for high-\nit\ case (4), and the lower emissivity bound for the low-\nit\ case (2), we adopt a plausible range of emissivities $ 0.47 < \epsilon < 0.72$ based on these results. 

It is unclear which of these cases best represents Pluto's real \nit-ice distribution in the south, but the results strongly suggest that Pluto's \nit\ ice has a bolometric emissivity below unity, with quite low values allowed. This is similar to conclusions drawn by \citet{stansberry1992tritonebal} regarding Triton's \nit\ ice, although we are arguably on much firmer footing regarding Pluto's energy balance because the LEISA spectra show where the \nit\ is on the surface (no such data were available from Voyager 2 for Triton). It is also consistent with preliminary results from the New Horizons REX radiometric measurements of the brightness temperature of Pluto's surface, which suggest that the directional, 4.2-cm emissivity (\ie\ non-bolometric) in Sputnik Planitia is approximately 0.7 (Linscott \etal, {\em submitted}). It is important to note, though, that the REX radiometric measurements were made at a very different wavelength from where the bulk of thermal emission happens.



\section{Discussion}\label{results}

We derived a map of the distribution of \nit\ ice on Pluto utilizing the LEISA spectra of the encounter hemisphere, merging lines of spectroscopic evidence from the encounter hemisphere and extrapolating the latitudinal \nit\ distribution there to other longitudes, to try and provide as realistic a map as possible. The resulting energy balance on Pluto's \nit\ ice suggests that the bolometric emissivity is likely in the range 0.4 -- 0.8, which is consistent with the range of estimated values for Triton based on a similar analysis \citep{stansberry1992tritonebal}. Below, we discuss the statistical nature of the \nit\ ice bolometric albedo, useful as an input to seasonal/climate models, and explore implications for the state of volatile transport on Pluto at the New Horizons epoch, and for the fate of Pluto's atmosphere at larger heliocentric distances.


\subsection{Latitudinal albedo trends}

\begin{figure} 
\centering
    \includegraphics[width=0.85\linewidth]{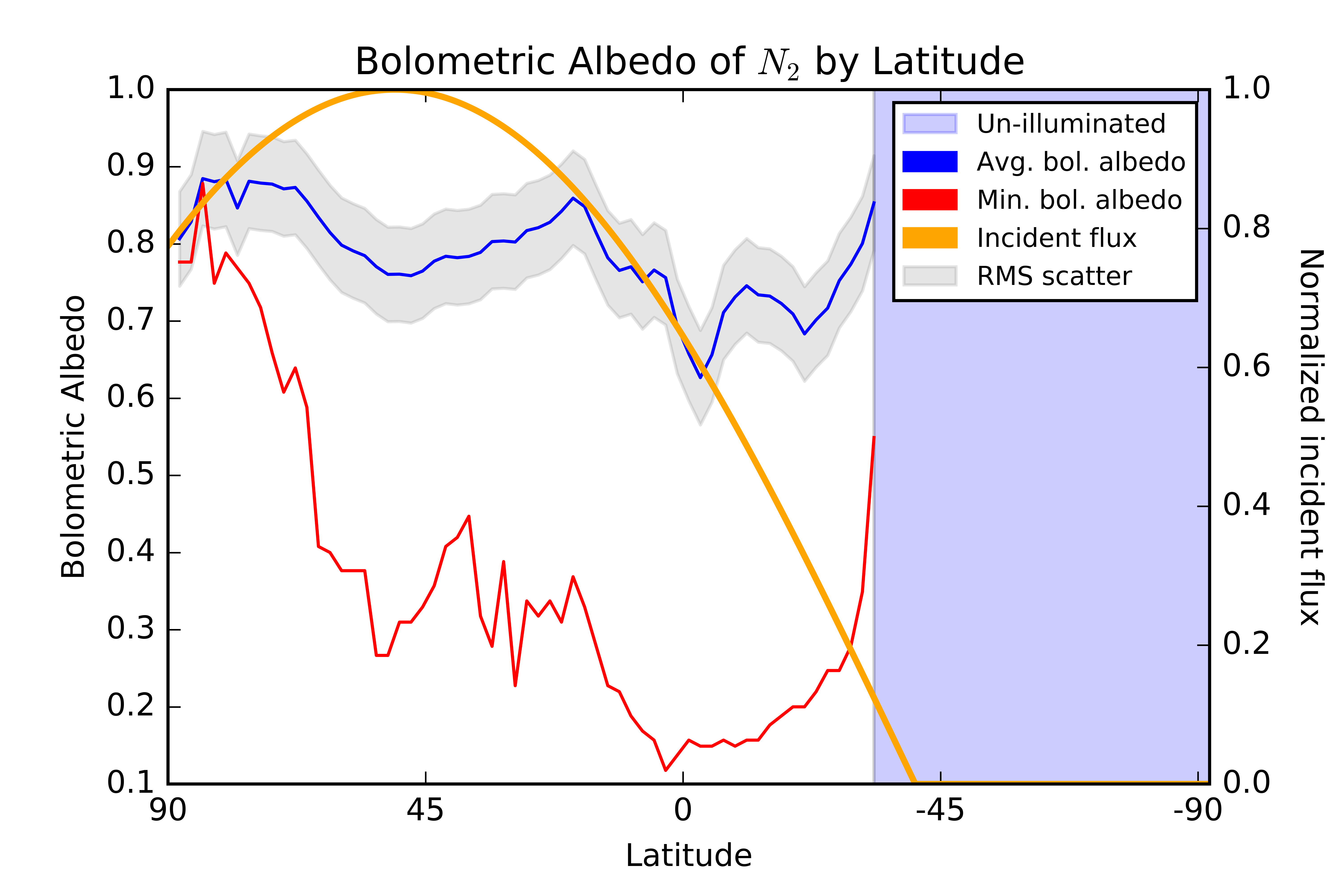}
\caption{Here we trace the average bolometric albedo of areas containing nitrogen for each latitude on Pluto. This line plot traces the variations in average albedo (blue), the RMS deviation about that average (gray shading), and the minimum albedo of the \nit\ ice at each latitude. The diurnally-averaged incident flux is also shown (orange), as calculated in Section \ref{energy}. The blue region at right represents the un-illuminated section of Pluto at the time of the New Horizons encounter.} 
\label{fig:avg-albedo}
\end{figure}

To further investigate the bolometric albedo of Pluto's \nit\ ice, we binned the albedo map (Figure~\ref{fig:albedo-map}) into sections of 1 degree of latitude. We then computed the average, root-mean-square (RMS) deviation, and minimum albedo of the \nit\ ice only (Figure \ref{fig:n2-coverage}) in each latitude bin, as shown in Figure \ref{fig:avg-albedo}. At latitudes north of +20\deg, the \nit\ ice albedo is relatively constant at about 0.8, with an RMS scatter of about 0.1. The albedo is slightly higher ($\simeq 0.85$) near +30\deg\, due to the very high albedo deposits in Sputnik Planitia. Albedo is also slightly higher above +70\deg, but there is relatively little \nit\ ice there in terms of the total area in that region or in terms of the fractional coverage (see Figure \ref{fig:n2-coverage}). The average bolometric albedo of the \nit\ ice near Pluto's equator falls to about 0.65, rising again to about 0.8 at the southern limit of the data. The \nit\ albedo minimum near the equator, and the lowest \nit\ ice albedos at each latitude (Fig.~\ref{fig:avg-albedo}), could result from pixels covering a patchwork of \nit\ ice and low albedo bare areas, a relatively transparent layer of \nit\ overlying low albedo substrate, or a combination of the two.

Areal mixing seems the more likely explanation for the apparently low \nit\ ice bolometric albedos near Pluto's equator (and the minimum albedos at all latitudes). The equator was strongly illuminated recently, in the years near equinox (1987), and low-albedo areas would have warmed significantly. Those warm surfaces would be unlikely to become sites of \nit\ condensation, particularly for enough condensation to yield a measurable \nit\ spectral feature.  If the low equatorial albedos are indeed the result of areal mixing of \nit\ and bare regions, then the actual albedo of \nit\ ice itself near the equator could be much higher than indicated in Figure~\ref{fig:avg-albedo}, and could easily be around 0.8, as it is further north and south. Further, this would imply that our energy balance calculations presented earlier somewhat overestimate the amount of insolation absorbed by the \nit\ ice. In turn that implies that the actual emissivity of the \nit\ ice would be somewhat lower than illustrated in Figure~\ref{fig:emissivity}. As can be seen in Figures~\ref{fig:n2-coverage} and \ref{fig:albedo-map}, the amount of low-albedo \nit\ near the equator is relatively small compared to the total inventory, so the overestimate of \tnit\ and \nit\ emissivity resulting from areal mixing of \nit\ ice and low albedo materials near the equator should be small. 

The average \nit\ ice bolometric albedo (and emissivity) on Pluto is of interest for climate and seasonal models. We find that the area-weighted average albedo, and RMS deviation, are 0.78 and 0.16, respectively. 
As can be seen in \ref{fig:albedo-map} and \ref{fig:avg-albedo}, there is a strong correlation between albedo and the presence of \nit\ ice. In particular, in \ref{fig:avg-albedo} the average albedo of the \nit\ is greater than 0.6 at all latitudes, and the RMS scatter of its albedo is about 0.1. There is a low-albedo tail to the albedo distribution of the \nit\ ice, but given the observations above, it seems likely that the tail is caused by mixing between \nit-ice and dark, \nit-free areas at the sub-pixel scale of the LEISA maps. To correct for the possibility such sub-pixel mixing we exclude some data as follows. Excluding the \nit\ albedo data from -20\deg -- +20\deg\ latitude, the average and RMS deviation are 0.79 and 0.14, respectively. Excluding albedos less than 0.6 at all latitudes, the average and RMS deviation are 0.83 (6\% larger than the global average given above) and 0.11, respectively. We note that the low-albedo tail for \nit\ ice in some areas will bias calculations of the energy balance somewhat (\ie\ to higher emissivities), but because they tend to occur for relatively few pixels, and mostly in areas with moderate to low insolation, that bias will be relatively small.

\subsection{Volatile Transport}\label{transport}

\begin{figure} 
    \centering
    \includegraphics[width=0.65\linewidth]{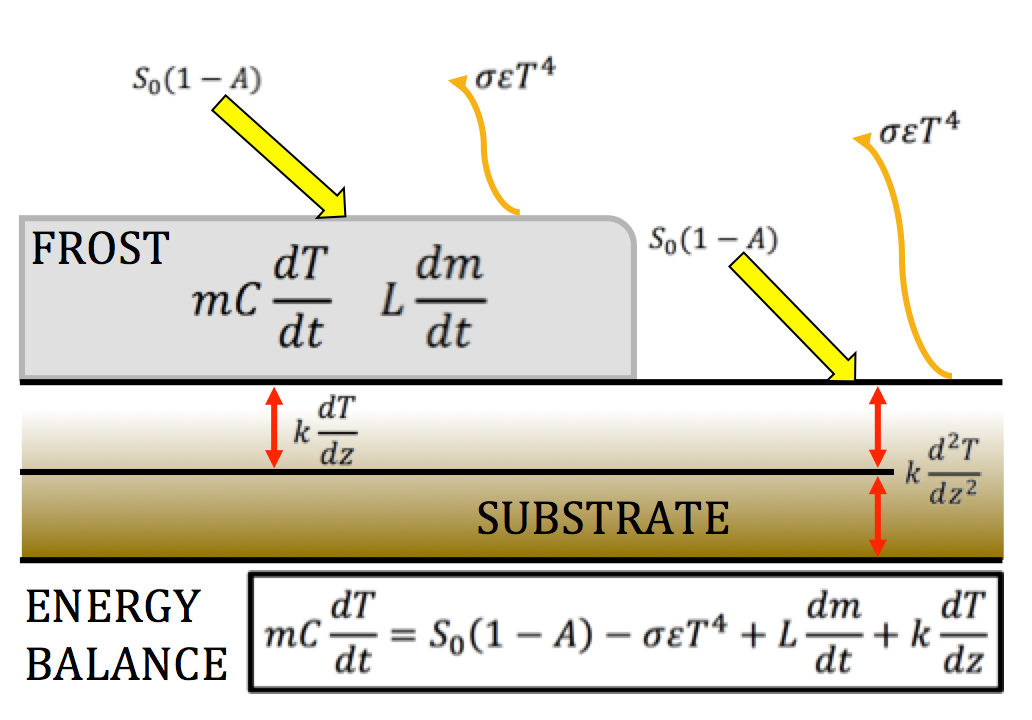}
    \caption{Visualization of energy balance on an icy surface in the presence of insolation ($S_0 (1-A)$, thermal radiation ($\epsilon \sigma T^4$), subsurface conduction ($k dT/dz$) and heat-capacity ($m C_p dT/dt$) and volatile transport ($L dm/dt$). Figure adapted from \citet{hansen1996seasonal}.}
    \label{fig:hansen-fig}
\end{figure}

As described in several previous studies (\eg\ \citealt{stern1988bright,hansen1996seasonal,spencer1997volatile,young2013vt3d}) Pluto's \nit\ (and other volatile) ices are constantly being transported across the surface in response to surpluses and deficits of insolation, carrying energy (as well as mass) away from high-insolation regions to low-insolation regions. Figure~\ref{fig:hansen-fig} illustrates the terms in the energy balance near Pluto's surface, including terms for insolation, re-radiation, sublimation/condensation, and subsurface conduction and heat capacity. For calculating the net effect of volatile transport on the distribution of \nit\ ice over seasonal and climatic timescales, the conduction and heat-capacity terms are very important, inducing a time-lag between the insolation cycle (driven by Pluto's obliquity and orbital eccentricity) and temperature variations and sublimation/condensation at the surface.

Our results were extended to provide a snapshot of the rates of sublimation and condensation of \nit\ ice at the time of the New Horizons encounter. For this investigation, we ignored the conduction and heat-capacity effects illustrated in Figure~\ref{fig:hansen-fig}. Including those effects would have been outside the scope of this work. Under the usual forward-modeling approach, this would have required an historical volatile ice distribution that would evolve, in a seasonal-transport model calculation, to match the observed distribution (and albedo patterns) described in Sections~\ref{sec:comp} and \ref{sec:albedo}. An alternative could be to take the \nit\ ice distribution we have derived, run a seasonal transport model backward in time, and seek values of the heat capacity and conductivity of the subsurface that matched historical constraints on the volatile distribution (\eg\ secular trends in atmospheric pressure, and/or the photometric and spectroscopic lightcurves). Either approach would be iterative, rely on oversimplified assumptions about the nature of Pluto's subsurface, be computationally expensive, and would not be guaranteed to provide a result that closely matches our \nit\ ice distribution.

In light of the above difficulties of using a full time-dependent volatile transport model to interpret our \nit\ ice distribution, we simplified the problem and considered the instantaneous state of volatile transport. Equation~\ref{balance-eq} gives the simplified form, neglecting the subsurface conduction and heat-capacity terms. Here $L$ is the latent heat of sublimation ($2.52\times10^9$\,erg/g), $\frac{dm}{dt}$ is the condensation mass flux, and the other terms are as in Equation~\ref{global-balance}. To solve this equation for the local mass flux, we set \tnit\ to 37.2\,K, consistent with the atmospheric pressure, and assume $\epsilon_{\nit} = 0.8$, based on our results above and consistent with the emissivity of Triton's \nit\ ice \citep{stansberry1992tritonebal}. By assuming that \tnit\ is a constant across Pluto's surface, we are implicitly requiring that the integral of the mass flux over all of the \nit\ ice is zero (\ie\ there is no net gain or loss of \nit\ mass due to the transport). In the results presented below we verified that the net calculated mass flux (\eg\ volatile inventory lost to the atmosphere) is very close to zero ($< 1\times 10^{-11}$\,g/cm$^2$/pixel). Elevation-driven transport rates (\citet{trafton2015departure}) are tiny relative to the insolation-driven rates on this timescale, and so were not included in these calculations. 

\begin{equation} 
    -L\frac{dm}{dt} = S_0(1-A)\cos\mu - \epsilon_{N_2}\sigma{T_{N_2}^4}
\label{balance-eq}
\end{equation}


\begin{figure} 
    \centering
    \begin{subfigure}{\linewidth}
    \centering
        \includegraphics[width=0.8\linewidth]{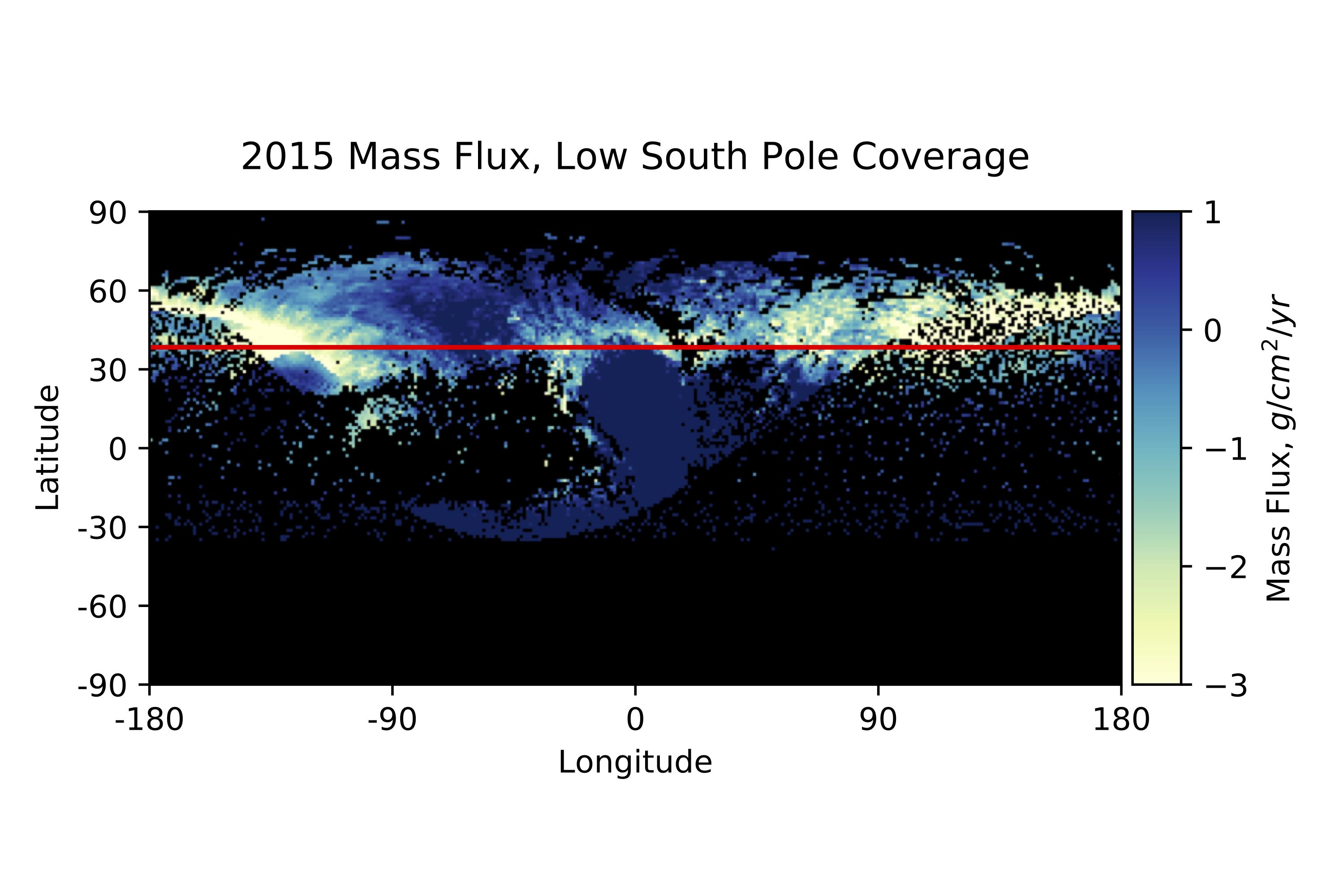}
        \vspace{-57pt}
    \end{subfigure}
    \begin{subfigure}{\linewidth}
    \centering
        \includegraphics[width=0.8\linewidth]{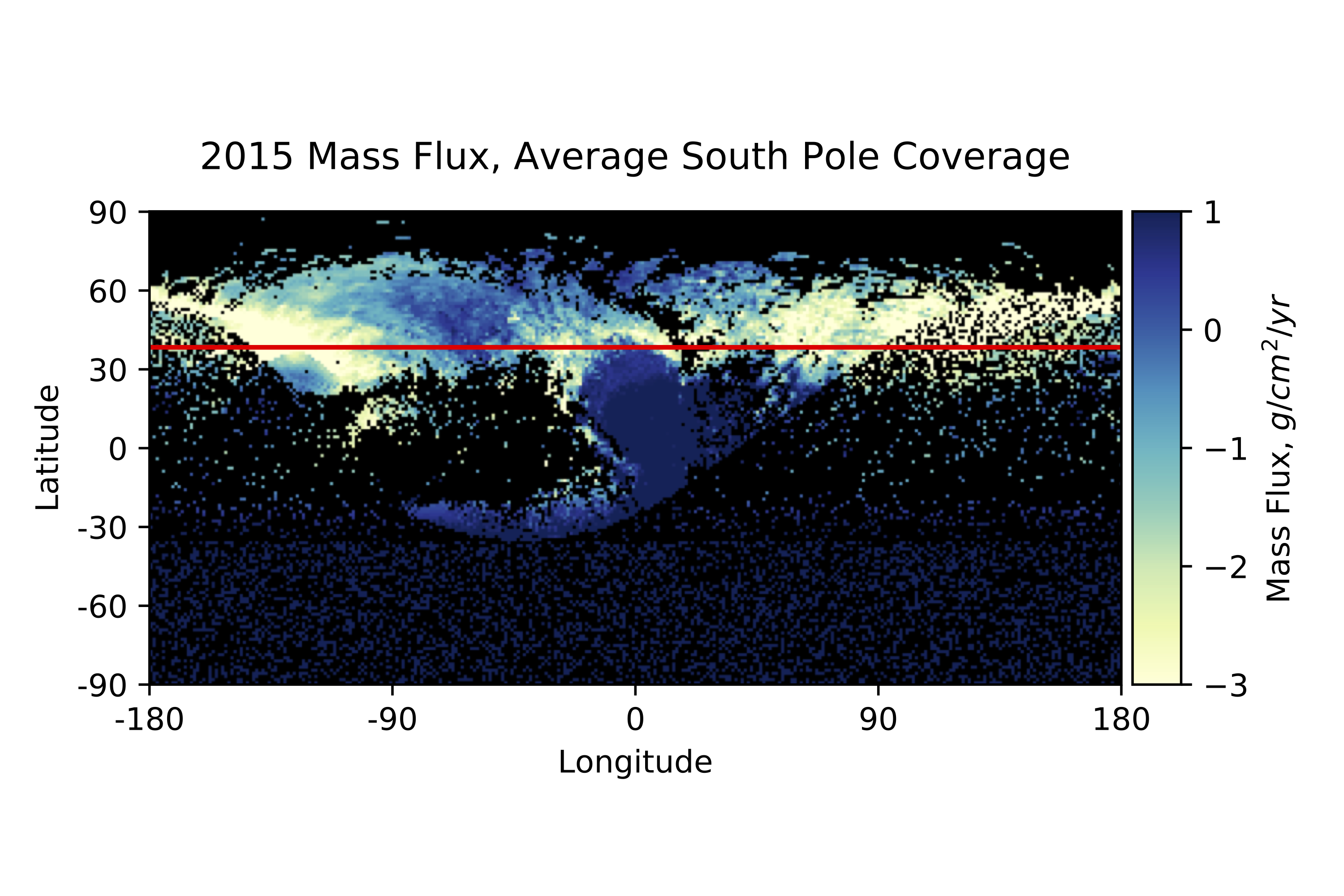}
        \vspace{-57pt}
    \end{subfigure}
    \begin{subfigure}{\linewidth}
    \centering
        \includegraphics[width=0.8\linewidth]{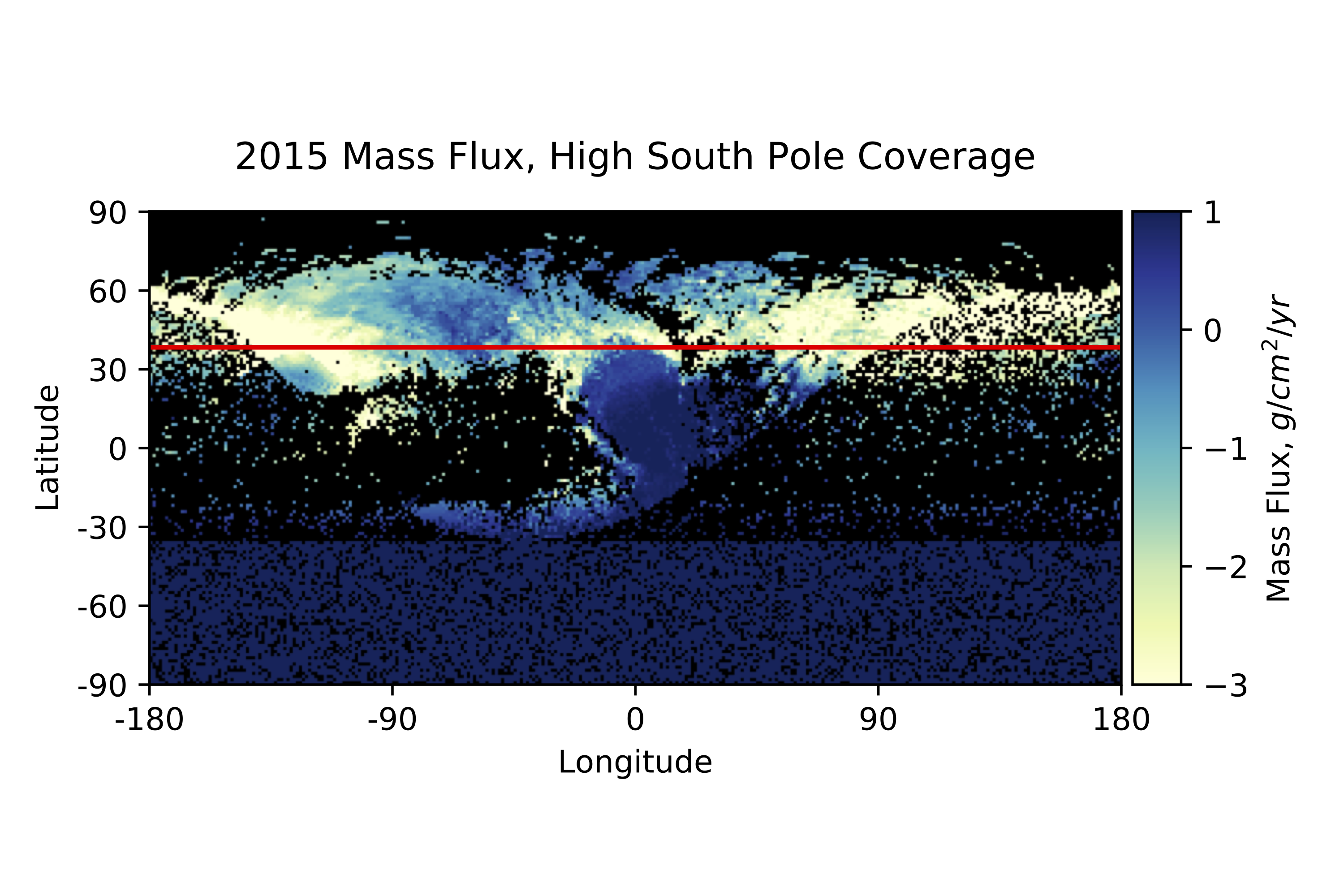}
        \vspace{-33pt}
    \end{subfigure}
    \caption{Mass fluxes of \nit\ ice condensation (positive) and sublimation (negative) based on our \nit\ ice distribution and energy balance calculations (Equation~\ref{balance-eq}). Red lines indicate the longitudinally-averaged sublimation/condensation boundary, \ie\ on average sublimation is occurring north of the line, and condensation to the south. Three cases are shown from the 5 we considered for the \nit\ ice distribution in the un-illuminated regions south of -38\deg. (Top) \nit\ ice coverage at the minimum value we derive for the encounter hemisphere; (Middle) coverage at the average encounter-hemisphere value; (Bottom) coverage at the maximum encounter-hemisphere value. The other two cases are no \nit\ ice coverage below -38\deg\ (which is very close to the minimum coverage case shown in the top panel, see Figure~\ref{fig:emissivity}), and completely covered in \nit. Details of the distribution of the \nit\ in the un-illuminated area are irrelevant for the mass fluxes; we illustrate them using random selection of pixels with the appropriate fractional coverage (note blue speckles). }
    \label{fig:mass-flux}
\end{figure}

The resulting mass fluxes of subliming and condensing nitrogen are shown in Figure~\ref{fig:mass-flux}. As mentioned in Section~\ref{sec:albedo}, the energy balance at the surface depends on the amount of \nit\ ice present south of -38\deg\ latitude, where there is no data. Also mentioned earlier, we considered 5 cases spanning the range of possible \nit\ coverage there, ranging from no-\nit\ to covered in \nit\ (see description of the cases in the discussion of Figure~\ref{fig:emissivity}). Three of those cases (representing coverage at the minimum, average, and maximum longitudinally-averaged coverage from Figure~\ref{fig:n2-coverage}) are represented in Figure~\ref{fig:mass-flux}. The patterns of condensation (positive mass fluxes) and sublimation are rather similar for all three cases shown. In general, there is transport from the higher-insolation areas in the north towards the south. Sputnik Planitia is, on average, experiencing condensation of \nit\ ice, except in the northernmost areas (we return to this topic below). The low-albedo \nit\ ice in western Cthulhu Regio is predicted to be actively sublimating (but this may not actually be the case if the \nit\ ice actually has a fairly high albedo and is segregated at the sub-pixel scale from even lower albedo Cthulhu-like material, as discussed earlier). According to \citet{bertrand2018nitrogen}, SP is currently at its minimum ice coverage in a sublimation-dominated epoch, so the sublimation we observe at the edges is expected. Their modeling work also suggests that the latitudinal band of \nit\ around 30$\deg$N is relatively stable, and may be explained by \met-rich deposits contributing to a high albedo in that region. 

For the low-\nit\ coverage case (top panel) the \nit\ condensation fluxes are significantly larger than for the average- and high-coverage cases (lower panels). This is because \nit\ ice is not present (and so is assumed unable to undergo condensation) in the un-illuminated area, and all of the mass sublimated in the illuminated region must also condense there. 
The calculated sublimation/condensation limit (where mass flux is 0) is 38.34{\deg}N, as shown in Figure~\ref{fig:mass-flux}. Because we are assuming the same surface pressure for all models, the location of the sublimation/condensation limit is independent of the assumed \nit-ice distribution in the un-illuminated regions. Instead, the magnitude of the sublimation rate changes to account for the changes in areas where condensation occurs.

The situations for the average- and high-\nit\ coverage cases (Figure~\ref{fig:mass-flux}, lower panels) are less likely to represent the volatile transport that was occurring at the encounter epoch. Of particular interest is the transition between sublimation and condensation within Sputnik Planitia basin. These results suggest that the northern-most part and the north-west border were undergoing weak sublimation, while condensation dominated over the larger southern area of Sputnik. The location of the sublimation/condensation transition agrees well with a contrast in the grain size of, and \met\ concentration in the \nit:\met\ component in the composition maps of \citet{protopapa2017pluto} and \citet{earle2017methane}, Figure~5. Our sublimating area correlates with their region (local to Sputnik) of smaller grain size and higher \met\ concentration in the \nit\ ice. The \met\ concentration correlation seems to be consistent with the preferential removal of \nit\ via sublimation of \nit:\met\ ice, as was suggested by \citet{stansberry1996metpatch}. The fact that the sublimation of \nit:\met\ seems to be correlated with smaller grain sizes for that component in northern Sputnik could be due to the surface becoming rougher, or grain boundaries widening, as sublimation proceeds, resulting in a more highly scattering material. However, an increase in scattering seems inconsistent with the slight albedo contrast between northern and mid- to southern Sputnik: bolometric albedo in our region of sublimation is $\simeq 0.94$, and is $\simeq 0.98$ further south in Sputnik. A similar gradient in normal reflectance is also seen \citep{buratti2017global}. This apparent discrepancy could be the result of the smaller grains being more forward scattering (and thus yielding a lower albedo in back-scatter) than the grains being formed further south, or (more likely) the concentration of lower-albedo contaminants in the \nit\ ice (\eg\ haze particles) as the ice sublimates.

\subsection{The fate of Pluto's atmosphere}\label{fate}

\citet{stansberry1996metpatch} explored the implications for low \nit\ ice emissivity for seasonal changes in Pluto's atmosphere. Specifically, they show that if there is an emissivity contrast between the solid $\alpha$ and $\beta$ phases of nitrogen, with the emissivity of \anit\ being lower, Pluto's atmospheric pressure would plateau at 4.8\mub, the vapor pressure at the $\alpha - \beta$ transition temperature (35.6\,K, \citet{fray2009vappress}). Hapke theory calculations of the bolometric emissivity of both phases of \nit\ \citep{stansberry1996emiss} show that \anit\ should in fact have a lower emissivity, with the emissivity contrast with \bnit\ being largest near the $\alpha - \beta$ transition temperature. The magnitude of the emissivity contrast is grain size dependent, but as long as the phase transition from \anit\ to \bnit\ does not spontaneously result in much larger grains of \anit, the contrast should be significant (a factor of $\simeq 3$ is typical in their calculations). The fact that \anit\ is slightly more dense than \bnit\ also suggests that as \anit\ forms from \bnit\ due to cooling, effective grain sizes may become smaller, \eg\ as grain boundaries open. The emissivity calculations neglected possible effects of \met\ and/or CO dissolved in the \nit\ ice. Either of those contaminants would raise the emissivity of the \nit\ because of their stronger far-IR absorption features. These contaminants also alter the transition temperature: \met\ lowers it, while CO raises it. Qualitatively, the effect of \met\ or CO dissolved in the \nit\ could be offset by assuming a reduced \nit\ grain size.

Our results here suggest that the emissivity of \bnit\ (the phase expected for \tnit$>35.6$\,K) on Pluto is indeed low. From the results in \citet{stansberry1996emiss}, the emissivity is consistent with effective grain sizes (for emitted wavelengths near 100\mum) of order 0.5\,cm, assuming pure \nit\ ice. The presence of \met\ diluted in the \nit\ ice will result in somewhat higher emissivity, so the effective grain size required to match our emissivity estimates would be somewhat smaller. At a grain size of 0.5 cm, the emissivity of \anit\ is expected to be a factor of 5 lower than for \bnit. As Pluto recedes from the Sun its \nit\ ice will cool to the $\alpha - \beta$ transition temperature, and \anit\ will begin to form. As that happens the effective globally averaged emissivity of the \nit\ ice will drop, causing its equilibrium temperature to stall at the phase-transition temperature. This effect could actually be stronger than discussed in \citet{stansberry1996emiss} because the phase transition is accompanied by a reduction in volume, which could result in smaller grains and/or stronger scattering at grain boundaries. Further, the solubility of \met\ is lower in \anit\ than in \bnit, so if dissolved \met\ is raising the emissivity relative to the earlier calculations, ex-solution of the \met\ during the phase change would increase the emissivity contrast between \anit\ and \bnit.

While not explicitly discussed in \citet{stansberry1999emissivity}, the simultaneous presence of the two solid phases (\anit\ and \bnit) in contact with the gas-phase atmosphere locks the \nit-ice/atmosphere system into a state with zero degrees of thermodynamic freedom until one of the phases is exhausted, so the state variables (\eg\ pressure and temperature) are also unable to change. Thus our finding that Pluto's current \nit-ice emissivity is low suggests that Pluto's atmospheric pressure may experience an extended period of stability at around 4.6\mub\ in the coming decades as \bnit\ converts to lower-emissivity \anit\ . 

There could be geologic expressions of the change from \anit\ to \bnit, although the situation is complex enough that one can only conjecture as to what they might be. It true that the density of \anit-ice is slightly higher than that of \bnit\ \citep{scott1976n2}; thus the cooling phase transition should result in smaller grains and perhaps larger or additional inter-grain spaces. This might result in an increase of albedo as the $\alpha$--$\beta$ phase transition occurs. The \anit\ will also tend to form where the \nit\ ice is `coolest', \eg\ on the winter hemisphere, nightside, and on topographic highs where the atmospheric pressure gradient favors sublimation over condensation, and the local equilibrium temperature can be slightly lower than the globally-defined \tnit. Detecting \anit\ ice in low- and no-illumination areas seems precluded for the most part, but a spectrometer on a fly-by or orbiting spacecraft, with sufficient sensitivity and resolving power, might be able to detect high-elevation day-side patches of \anit, particularly if it included the much stronger 4.3\mum\ band of \nit.

\section{Conclusions}


In this work, we created a new map of the \nit\ ice distribution on Pluto, synthesized from previously published composition maps of the New Horizons encounter hemisphere, and extrapolated to Pluto's entire surface. Because it is based in part on the detection of bands of \met\ dissolved in the \nit\ ice, this has resulted in a more complete description of the geographic distribution of the \nit\ ice than possible using its elusive 2.15-\mum\ absorption feature alone. We used the map to compute the latitudinal and altitude dependent distribution of the \nit, and its average albedo. We then examined the global energy balance of the \nit\ ice and used the measured surface pressure to constrain its bolometric emissivity. 

We found that global-scale topography is anti-correlated with strong \nit-ice absorption band depths, indicating large grains of that ice particularly in (but not limited to) the huge Sputnik Planitia basin. This was also reported soon after the 2015 encounter (\eg\ \citet{grundy2016surface}), and our new map of the \nit\ distribution does not change that conclusion. This confirms pre-encounter predictions of long-term down-hill transport of \nit (\eg\ \citet{trafton2015departure}), and is also predicted by much more complex global circulation models (\eg\  \citet{bertrand2016atmotopo}). However there are also higher-altitude deposits of \nit\ ice in a broad band between 30\deg N and 60\deg N latitude, typically at altitudes of 0 to 0.5 km and having generally weaker 2.15-\mum\ band depths than observed in Sputnik. We also provided a lower bound on the total amount of \nit\ on Pluto, on the order of $10^5$ km$^3$, illustrating that a large fraction of the nitrogen is contained within Sputnik Planitia.

The latitudinally averaged bolometric albedo of Pluto's \nit\ ice ranges from 0.65 to 0.95. Excluding albedos lower than 0.6, which probably result from areal mixing of \nit\ and darker materials at the sub-pixel scale, the global average bolometric albedo is 0.83, with an RMS scatter of 0.11, similar to that inferred for Triton. In the case of Pluto, the value is better substantiated because it is based on the actual distribution of the \nit\ ice inferred from the LEISA spectral maps, whereas for Triton no such composition maps were available. On the other hand, Triton's photometric function was better constrained than Pluto's. 

Pluto's \nit\ ice probably has a low emissivity, and the actual value depends on assumed \nit\ ice fractional coverage in the un-illuminated areas south of -38\deg, but is probably in the range 0.47 -- 0.72, similar to Triton. These low \bnit-ice emissivities are consistent with the results of \citet{stansberry1996emiss}, which predict even lower emissivities for \anit. As described in \citet{stansberry1999emissivity}, once Pluto's \nit\ ice reaches the $\alpha$--$\beta$ transition temperature of 35.6\,K, vapor pressure balance between the two solid phases and Pluto's atmosphere will tend to stabilize the atmospheric pressure at 4.6\mub\ until all of the \nit\ ice in contact with the atmosphere converts to one form or the other. This effect could result in Pluto having a significant atmosphere even at aphelion, and could be tested in coming decades by using stellar occultations to measure the atmospheric pressure.

Calculated patterns of volatile transport suggest global north-to-south redistribution, with Sputnik being a northward extension of the region of condensation, caused by its high albedo. Within Sputnik, sublimation is probably occurring in the north and northwest, condensation everywhere else, consistent with details of the spectral fits. These results are also consistent with the post-New Horizons global climate models of \citet{forget2017post}. Typical sublimation and deposition rates are approximately 1 g\,cm$^{-2}$\,yr$^{-1}$, with sublimation being more geographically concentrated and therefore somewhat stronger. The low emissivity we find for the \nit\ ice tends to slightly enhance rates of sublimation, and decrease rates of condensation, relative to what they would be were the emissivity higher.

There are natural extensions of our energy-balance analysis that are currently limited by the available calibrated data products.For example, LEISA maps for the \nit-ice distribution on the non-encounter hemisphere would improve the accuracy of our results, and are being worked on at this time. While the phase function for the \nit-ice dominated areas also introduces a systematic uncertainty in the phase integral used to calculate the bolometric albedo, such an improvement is unlikely given the limited phase-angle coverage available from New Horizons. Over the long term, measurements of Pluto's atmospheric pressure via stellar occultation will test the validity of the prediction that the transition of \bnit\ to \anit\ will result in a stable atmosphere even as Pluto approaches aphelion.


\section*{Acknowledgements}

The authors would like to thank NASA for supporting the New Horizons project, and the New Horizons team for making this mission and the results of the flyby possible. This work was supported in part by Space Telescope Science Institute, as a part of their 2017 Summer Astronomy Student Program (SASP), and also in part by NASA Solar System Workings grant NNX15AH35G. Calculations and generation of figures made use of the {\em astropy} and {\em ISIS3} software packages.



\section*{References}

\bibliographystyle{elsarticle-num}
\bibliography{pluto-bib}

\end{document}